\documentclass[aps,twocolumn,11pt,floatfix,altaffilletter,superscriptaddress,preprintnumbers,tightenlines,showpacs,showkeys,notitlepage,nofootinbib]{revtex4-2}
\usepackage[hidelinks,colorlinks,pdfusetitle]{hyperref}
\usepackage{geometry}
\usepackage[normalem]{ulem}
\usepackage{amsmath,amssymb}
\usepackage{relsize}
\usepackage{mathtools}
\usepackage{verbatim}
\usepackage{titlesec}               % modify title and section heading format
\usepackage{epsfig}
\usepackage{graphicx}               % Standard graphics package
\usepackage{url}
\usepackage{color}
\usepackage{multirow}
\usepackage{placeins}
\usepackage[dvipsnames]{xcolor}

\definecolor{kleingrothaus}{HTML}{009FC3}
\hypersetup{colorlinks=true,allcolors=kleingrothaus}

\usepackage{epstopdf}
\usepackage{siunitx}
\usepackage{fontawesome}
\usepackage{enumitem}
\usepackage[capitalize]{cleveref}
\usepackage{lipsum}
\usepackage{gensymb}
\usepackage{nicematrix}
\usepackage{booktabs}               % nicer looking tables
\usepackage{xspace}                 % smart space after command
\usepackage{pifont}
\usepackage{marvosym }%lightning symbol
\usepackage{tikz-feynman}
\usepackage{tikz} % to draw Feynman diagrams manually
\usetikzlibrary{shapes,arrows}
\usetikzlibrary{decorations.pathmorphing,decorations.markings}
\usetikzlibrary{snakes}
\usetikzlibrary{positioning,arrows.meta}
\tikzfeynmanset{ fermion/.style = {
   decoration={
     markings,
     mark=at position 0.5
          with {\arrow[xshift=2mm]{Stealth[black,width=2mm,length=3mm]}}
     },
   postaction=decorate}
}

\usepackage{orcidlink}  % to add ORCID 
\usepackage{upgreek}
\allowdisplaybreaks

\usepackage{bbm}
\usepackage{slashed}

\makeatletter

\renewcommand{\p@subsection}{}
\makeatother

\titleformat*{\section}{\centering\bfseries\uppercase}
\titlelabel{\thetitle\quad}
\titleformat*{\paragraph}{\bfseries}
\titlespacing*{\paragraph}{0pt}{3.25ex plus 1ex minus .2ex}{1em}

\makeatletter
\def\l@subsubsection#1#2{}
\makeatother

\DeclarePairedDelimiter\chevron{\langle}{\rangle}

\newlist{datalist}{description}{1}
\setlist[datalist]{font=\normalfont\itshape, style=multiline,
  leftmargin=4.9em, labelsep=0.5em, nosep, topsep=1pt, itemsep=1pt}

\newcommand{\mbb}{\ensuremath{m_{\beta\beta}}}
\newcommand{\VmA}{\ensuremath{|\upvarepsilon_{V-A}^{V+A}|}}
\newcommand{\VpA}{\ensuremath{|\upvarepsilon_{V+A}^{V+A}|}}
\newcommand{\SP}{\ensuremath{|\upvarepsilon_{S\pm P}^{S+P}|}}
\newcommand{\TR}{\ensuremath{|\upvarepsilon_{T_R}^{T_R}|}}
\newcommand{\pec}{\ensuremath{\mathrm{EC}\beta^+}}
\newcommand{\pp}{\ensuremath{\beta^+\beta^+}}
\newcommand{\mm}{\ensuremath{\beta^-\beta^-}}
\newcommand{\zvbb}{\ensuremath{0\nu\beta\beta}}
\newcommand{\bmu}{\textcolor{cVLL}{\ensuremath{V_{LL}}}}
\newcommand{\beeta}{\textcolor{cVLR}{\ensuremath{V_{LR}}}}
\newcommand{\blam}{\textcolor{cVRR}{\ensuremath{V_{RR}}}}
\newcommand{\bsig}{\textcolor{cS}{\ensuremath{S}}}
\newcommand{\btau}{\textcolor{cT}{\ensuremath{T}}}

\definecolor{cVLL}{RGB}{0,114,178}   % V_LL  (blue)
\definecolor{cVLR}{RGB}{0,158,115}   % V_LR  (green)
\definecolor{cVRR}{RGB}{213,94,0}    % V_RR  (orange)
\definecolor{cT}{RGB}{204,121,167}   % T     (pink)
\definecolor{cS}{RGB}{230,159,0}     % S     (gold)

\newcommand{\cmu}{\textcolor{cVLL}{\bmu}}
\newcommand{\ceta}{\textcolor{cVLR}{\beeta}}
\newcommand{\clam}{\textcolor{cVRR}{\blam}}
\newcommand{\ctau}{\textcolor{cT}{\btau}}
\newcommand{\csig}{\textcolor{cS}{\bsig}}

\definecolor{rhopos}{HTML}{B2182B}   % positive rho — red
\definecolor{rhoneg}{HTML}{2166AC}   % negative rho — blue
\begin{document}
%=============================================================================
% \title{\textcolor{kleingrothaus}{Competing neutrinoless decay modes\\ as a probe of lepton number violation}}
\title{\textcolor{kleingrothaus}{Pinpointing the Mechanism of Neutrinoless Weak Decays with Positrons}}
\author{Julia Harz \orcidlink{0000-0002-8362-4083}}
\affiliation{PRISMA$^{++}$ Cluster of Excellence \& Mainz Institute for Theoretical Physics, \\ FB 08 - Physics,
Mathematics and Computer Science,\\
Johannes Gutenberg-Universit{\"a}t Mainz,
55099 Mainz, Germany}
\author{George A. Parker \orcidlink{0009-0000-1836-8696}}
\affiliation{PRISMA$^{++}$ Cluster of Excellence \& Mainz Institute for Theoretical Physics, \\ FB 08 - Physics,
Mathematics and Computer Science,\\
Johannes Gutenberg-Universit{\"a}t Mainz,
55099 Mainz, Germany}
%=============================================================================
\begin{abstract}

Neutrinoless double beta decay is the flagship laboratory probe of a Majorana contribution to the neutrino mass. However, besides the standard mass mechanism other higher-dimensional lepton number-violating interactions can enter, or even dominate, this process. The corresponding positron-emitting neutrinoless modes, such as electron capture or double-positron emission, have long been considered out of reach experimentally, due to their naturally smaller rates. Recently, innovative detector technologies as used in the proposed NuDoubt$^{++}$ experiment are changing the game. In this work, we explore how a positron- and electron-mode detector, can be complementary in the search for new physics. Assuming an observation of neutrinoless double beta decay, we predict the expected discovery half-life for the positron-modes. Using half-life \emph{ratios}, especially between the double beta and electron capture modes, we demonstrate how underlying long-range interactions can be distinguished, in particular to identify a purely right-handed leptonic current.

\end{abstract}
\maketitle
%=============================================================================
\section{Introduction}
\label{sec:intro}
%=============================================================================
The observation of neutrinoless double beta ($\zvbb$) decay would establish lepton number violation (LNV) and point towards a Majorana nature of the neutrino, making it a central target of the current experimental program in fundamental physics~\cite{Roadmap}. The most stringent bound to date comes from KamLAND-Zen, $T^{0\nu}_{1/2}[^{136}\mathrm{Xe}] > 3.8\times 10^{26}$~yr at
90\%~C.L.~\cite{KamLAND-Zen:2024eml}. Future $\zvbb$ experiments such as KamLAND2-Zen or LEGEND aim to reach lifetimes $T_{1/2}^{\zvbb}$ of order $10^{27}$ yr ($^{136}$Xe) and $10^{28}$ yr ($^{76}$Ge), respectively \cite{Shirai:2017jyz,LEGEND:2021bnm}.
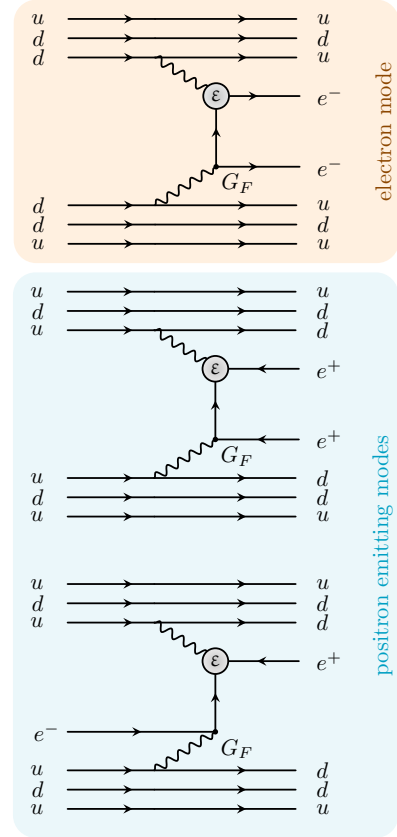
\begin{figure}[t]
\centering
% ================= electron mode =================
\scalebox{0.85}{\begin{tikzpicture}[x=1cm,y=1cm,>=stealth,line cap=round]
\tikzset{
qleft/.style={thick,postaction={decorate},decoration={markings,mark=at position 0.75 with {\arrow{>}}}},
qright/.style={thick,postaction={decorate},decoration={markings,mark=at position 0.65 with {\arrow{>}}}},
wline/.style={thick,decorate,decoration={snake,amplitude=1.8pt,segment length=6pt}},
nuline/.style={thick,postaction={decorate},decoration={markings,mark=at position 0.55 with {\arrow{>}}}},
eline/.style={thick,postaction={decorate},decoration={markings,mark=at position 0.5 with {\arrow{>}}}},
epsblob/.style={thick,shade,shading=radial,inner color=gray!45,outer color=gray!18,draw=black}
}
\fill[orange!12,rounded corners=10pt] (-4.05,-2.05) rectangle (1.95,2.05);
\node[rotate=90,font=\small,text=orange!55!black] at (1.70,0) {electron mode};
\coordinate (Tq) at (-1.85,1.15);
\coordinate (Bq) at (-1.85,-1.15);
\coordinate (Tv) at (-0.90,0.55);
\coordinate (Bv) at (-0.90,-0.55);
\draw[qleft] (-3.2,1.75)--(-1.85,1.75); \draw[qright] (-1.85,1.75)--(0.35,1.75);
\draw[qleft] (-3.2,1.45)--(-1.85,1.45); \draw[qright] (-1.85,1.45)--(0.35,1.45);
\draw[qleft] (-3.2,1.15)--(Tq); \draw[qright] (Tq)--(0.35,1.15);
\draw[qleft] (-3.2,-1.15)--(Bq); \draw[qright] (Bq)--(0.35,-1.15);
\draw[qleft] (-3.2,-1.45)--(-1.85,-1.45); \draw[qright] (-1.85,-1.45)--(0.35,-1.45);
\draw[qleft] (-3.2,-1.75)--(-1.85,-1.75); \draw[qright] (-1.85,-1.75)--(0.35,-1.75);
\node[anchor=east] at (-3.45,1.75) {$u$}; \node[anchor=east] at (-3.45,1.45) {$d$}; \node[anchor=east] at (-3.45,1.15) {$d$};
\node[anchor=west] at (0.55,1.75) {$u$}; \node[anchor=west] at (0.55,1.45) {$d$}; \node[anchor=west] at (0.55,1.15) {$u$};
\node[anchor=east] at (-3.45,-1.15) {$d$}; \node[anchor=east] at (-3.45,-1.45) {$d$}; \node[anchor=east] at (-3.45,-1.75) {$u$};
\node[anchor=west] at (0.55,-1.15) {$u$}; \node[anchor=west] at (0.55,-1.45) {$d$}; \node[anchor=west] at (0.55,-1.75) {$u$};
\draw[wline] (Tq)--(Tv); \draw[wline] (Bq)--(Bv);
\fill (Bv) circle (1.2pt);
\node[anchor=north west] at (-0.94,-0.54) {$G_F$};
\draw[nuline] (Bv)--(-0.90,0.55);
% \draw[nuline] (Tv)--(-0.90,0.06);
% \draw[nuline] (Bv)--(-0.90,-0.06);
% \draw[thick] (-1.02,-0.12)--(-0.78,0.12); \draw[thick] (-1.02,0.12)--(-0.78,-0.12);
% \node[anchor=east] at (-1.08,0) {$\mbb$};
\draw[eline] (Tv)--(0.40,0.55); \draw[eline] (Bv)--(0.40,-0.55);
\node[anchor=west] at (0.55,0.55) {$e^{-}$}; \node[anchor=west] at (0.55,-0.55) {$e^{-}$};
\draw[epsblob] (Tv) circle (0.20); \node[font=\small] at (Tv) {$\upvarepsilon$};
\end{tikzpicture}}\\[0.25em]
% ================= positron emitting mode =================
\scalebox{0.85}{\begin{tikzpicture}[x=1cm,y=1cm,>=stealth,line cap=round]
\tikzset{
qleft/.style={thick,postaction={decorate},decoration={markings,mark=at position 0.75 with {\arrow{>}}}},
qright/.style={thick,postaction={decorate},decoration={markings,mark=at position 0.65 with {\arrow{>}}}},
wline/.style={thick,decorate,decoration={snake,amplitude=1.8pt,segment length=6pt}},
nuline/.style={thick,postaction={decorate},decoration={markings,mark=at position 0.55 with {\arrow{>}}}},
eline/.style={thick,postaction={decorate},decoration={markings,mark=at position 0.5 with {\arrow{>}}}},
pline/.style={thick,postaction={decorate},decoration={markings,mark=at position 0.5 with {\arrow{>}}}},
epsblob/.style={thick,shade,shading=radial,inner color=gray!45,outer color=gray!18,draw=black}
}
\fill[kleingrothaus!8,rounded corners=10pt] (-4.05,-6.75) rectangle (1.95,2.05);
\node[rotate=90,font=\small,text=kleingrothaus] at (1.70,-2.35) {positron emitting modes};
% ---------- beta+ beta+ ----------
\begin{scope}
\coordinate (Tq) at (-1.85,1.15);
\coordinate (Bq) at (-1.85,-1.15);
\coordinate (Tv) at (-0.90,0.55);
\coordinate (Bv) at (-0.90,-0.55);
\draw[qleft] (-3.2,1.75)--(-1.85,1.75); \draw[qright] (-1.85,1.75)--(0.35,1.75);
\draw[qleft] (-3.2,1.45)--(-1.85,1.45); \draw[qright] (-1.85,1.45)--(0.35,1.45);
\draw[qleft] (-3.2,1.15)--(Tq); \draw[qright] (Tq)--(0.35,1.15);
\draw[qleft] (-3.2,-1.15)--(Bq); \draw[qright] (Bq)--(0.35,-1.15);
\draw[qleft] (-3.2,-1.45)--(-1.85,-1.45); \draw[qright] (-1.85,-1.45)--(0.35,-1.45);
\draw[qleft] (-3.2,-1.75)--(-1.85,-1.75); \draw[qright] (-1.85,-1.75)--(0.35,-1.75);
\node[anchor=east] at (-3.45,1.75) {$u$}; \node[anchor=east] at (-3.45,1.45) {$d$}; \node[anchor=east] at (-3.45,1.15) {$u$};
\node[anchor=west] at (0.55,1.75) {$u$}; \node[anchor=west] at (0.55,1.45) {$d$}; \node[anchor=west] at (0.55,1.15) {$d$};
\node[anchor=east] at (-3.45,-1.15) {$u$}; \node[anchor=east] at (-3.45,-1.45) {$d$}; \node[anchor=east] at (-3.45,-1.75) {$u$};
\node[anchor=west] at (0.55,-1.15) {$d$}; \node[anchor=west] at (0.55,-1.45) {$d$}; \node[anchor=west] at (0.55,-1.75) {$u$};
\draw[wline] (Tq)--(Tv); \draw[wline] (Bq)--(Bv);
\fill (Bv) circle (1.2pt);
\node[anchor=north west] at (-0.94,-0.54) {$G_F$};
\draw[nuline] (Bv)--(-0.90,0.55);
% \draw[nuline] (Tv)--(-0.90,0.06);
% \draw[nuline] (Bv)--(-0.90,-0.06);
% \draw[thick] (-1.02,-0.12)--(-0.78,0.12); \draw[thick] (-1.02,0.12)--(-0.78,-0.12);
% \node[anchor=east] at (-1.08,0) {$\mbb$};
\draw[pline] (0.40,0.55)--(Tv); \draw[pline] (0.40,-0.55)--(Bv);
\node[anchor=west] at (0.55,0.55) {$e^{+}$}; \node[anchor=west] at (0.55,-0.55) {$e^{+}$};
\draw[epsblob] (Tv) circle (0.20); \node[font=\small] at (Tv) {$\upvarepsilon$};
\end{scope}
% ---------- EC beta+ ----------
\begin{scope}[yshift=-4.55cm]
\coordinate (Tq) at (-1.85,1.15);
\coordinate (Bq) at (-1.85,-1.15);
\coordinate (Tv) at (-0.90,0.55);
\coordinate (Bv) at (-0.90,-0.55);
\draw[qleft] (-3.2,1.75)--(-1.85,1.75); \draw[qright] (-1.85,1.75)--(0.35,1.75);
\draw[qleft] (-3.2,1.45)--(-1.85,1.45); \draw[qright] (-1.85,1.45)--(0.35,1.45);
\draw[qleft] (-3.2,1.15)--(Tq); \draw[qright] (Tq)--(0.35,1.15);
\draw[qleft] (-3.2,-1.15)--(Bq); \draw[qright] (Bq)--(0.35,-1.15);
\draw[qleft] (-3.2,-1.45)--(-1.85,-1.45); \draw[qright] (-1.85,-1.45)--(0.35,-1.45);
\draw[qleft] (-3.2,-1.75)--(-1.85,-1.75); \draw[qright] (-1.85,-1.75)--(0.35,-1.75);
\node[anchor=east] at (-3.45,1.75) {$u$}; \node[anchor=east] at (-3.45,1.45) {$d$}; \node[anchor=east] at (-3.45,1.15) {$u$};
\node[anchor=west] at (0.55,1.75) {$u$}; \node[anchor=west] at (0.55,1.45) {$d$}; \node[anchor=west] at (0.55,1.15) {$d$};
\node[anchor=east] at (-3.45,-1.15) {$u$}; \node[anchor=east] at (-3.45,-1.45) {$d$}; \node[anchor=east] at (-3.45,-1.75) {$u$};
\node[anchor=west] at (0.55,-1.15) {$d$}; \node[anchor=west] at (0.55,-1.45) {$d$}; \node[anchor=west] at (0.55,-1.75) {$u$};
\draw[wline] (Tq)--(Tv); \draw[wline] (Bq)--(Bv);
\fill (Bv) circle (1.2pt);
\node[anchor=north west] at (-0.94,-0.54) {$G_F$};
% \draw[nuline] (Tv)--(-0.90,0.06);
% \draw[nuline] (Bv)--(-0.90,0.06);
\draw[nuline] (Bv)--(-0.90,0.55);
% \draw[thick] (-1.02,-0.12)--(-0.78,0.12); \draw[thick] (-1.02,0.12)--(-0.78,-0.12);
% \node[anchor=east] at (-1.08,0) {$\mbb$};
\draw[pline] (0.40,0.55)--(Tv); \draw[eline] (-3.20,-0.55)--(Bv);
\node[anchor=west] at (0.55,0.55) {$e^{+}$}; \node[anchor=east] at (-3.20,-0.55) {$e^{-}$};
\draw[epsblob] (Tv) circle (0.20); \node[font=\small] at (Tv) {$\upvarepsilon$};
\end{scope}
\end{tikzpicture}}
\caption{Long-range contributions to the three neutrinoless modes considered in this work ($\zvbb$, $0\nu\pp$, and $0\nu\pec$). In each diagram, one vertex carries the Standard Model $V-A$ interaction ($G_F$) and the other a general effective coupling $\upvarepsilon$, drawn as a shaded blob. In each diagram, this interaction could be replaced with two weak vertices, and a neutrino mass insertion.}
\vspace{-0.5cm}
\label{fig:diag-bbm}
\label{fig:diag-pos}
\end{figure}
In the standard picture, this process ($\zvbb$) converts two neutrons to two protons via the exchange of a light Majorana neutrino, parametrized by the effective mass $\mbb$. 

However, analogous neutrinoless decay modes also exist on the proton-rich side of the valley of stability, in which two protons are converted into two neutrons. Such a transition can occur via three distinct channels: double-positron emission ($0\nu\pp$), positron-emitting electron capture ($0\nu\pec$), or double-electron capture ($0\nu$ECEC). These positron modes have long been considered experimentally inaccessible due to their phase-space suppression. Recently this assessment has been changing; the proposed NuDoubt$^{++}$ experiment~\cite{NuDoubt:2024jax}, based on hybrid opaque loaded scintillator techniques, is targeting observation of positron-emitting double beta decay with $^{78}$Kr (and then $^{106}$Cd and $^{124}$Xe in later phases) with a projected $0\nu\pp$ decay half-life sensitivity of order $10^{24}$~yr. Hence, the question is not anymore whether the positron modes can compete with $0\nu\mm$ searches with respect to their corresponding half-life reach---they cannot, for a long time---but whether they carry complementary information and at what exposure that information becomes relevant. 

Discriminating between different long-range LNV physics requires additional measurements or observables \cite{Deppisch2012,Graf2022}. Proposals in this direction include half-life ratios of different isotopes~\cite{Deppisch:2006hb,Bergstrom2011}, angular and energy distributions of the electrons~\cite{Doi:1985dx,Graf2022}, or the different decay modes, $0\nu\pp$  and $0\nu\pec$~\cite{Doi:1992dm,Hirsch1994,Kotila2013}, which is the focus of this work.

By modifying the current formalism for $0\nu\mm$, in this work,
we first study the single-current sensitivities (\cref{sec:single}), and then answer three related questions on the promise of considering these positron-emitting modes. 
\begin{enumerate}[label=\Roman*.]
\item At what half-life sensitivity does NuDoubt$^{++}$ provide complementary information with respect to $0\nu\mm$ decay? (\cref{sec:nudoubt})
\item How well do half-life \emph{ratios} across the three decay modes and for different isotopes discriminate between different long-range LNV physics? (\cref{sec:DPratios})
\item How robust is the analysis in case of several LNV effective couplings present at a time, and can half-life ratios still be used to discriminate between new interactions? (\cref{sec:interference}) 
\end{enumerate}
The remaining part of the paper is structured as follows: In \cref{sec:longrange} we explain the details of the long-range physics studied in this work, and in \cref{sec:rate} and \cref{sec:crossing} we show how this enters the expression for the half-life for all three modes. In \cref{sec:inputs}, we discuss the nuclear inputs used in this study and present our results in \cref{sec:single} to \cref{sec:interference}. Finally, we conclude in \cref{sec:conclusions}.

\textbf{Note added:} While this work was being completed, \cite{GKS26} appeared, presenting an independent analysis of the NuDoubt$^{++}$ positron-emitting isotopes, and an interpretation in terms of dimension-seven SMEFT operators. Our works have different focuses, but where our analyses overlap, we find agreement. The complementary relation between the two works is discussed in \cref{sec:gks}. We also update our positron-emitting nuclear inputs from~\cite{Barea2015} to the newer version in~\cite{GKS26}.

\section{Theory of Neutrinoless Double Weak Decay}
\subsection{Long-Range Currents}
\label{sec:longrange}

For all three modes, the decay rate factorizes, \emph{schematically}, as $[T^{0\nu}_{1/2}]^{-1}\sim G\,|M|^{2}\,|\upvarepsilon|^{2}$: a phase-space factor (PSF) $G$ encoding the kinematics of the emitted or captured leptons, a nuclear matrix element (NME) $M$ encoding the structure of the decaying nucleus, and the unknown coupling, $\upvarepsilon$. The latter effectively describes the new physics involved, and we aim to study in the following how the positron-emitting modes can help to distinguish different $\upvarepsilon$'s \cite{Hirsch1994}.

In addition to the standard light neutrino exchange, other $\Delta L = 2$ interactions can generate neutrinoless decay modes such as $0\nu\mm$, $0\nu\pp$, and $0\nu\pec$. Long-range contributions (\cref{fig:diag-bbm}) can be described by two vertices that are pointlike at the Fermi scale with a light neutrino being exchanged between them. The general effective Lagrangian reads~\cite{Pas1999,Pas2001,Deppisch2012}
\begin{equation}
\mathcal{L} = \frac{G_F}{\sqrt{2}}\Bigl\{\, j^{\mu}_{V-A}\,J^{\dagger}_{V-A,\mu}
   \;+\; {\sum_{\alpha,\beta}}'\, \upvarepsilon^{\alpha}_{\beta}\, j_{\beta}\,J^{\dagger}_{\alpha} \Bigr\},
\label{eq:Leff}
\end{equation}
built from hadronic and leptonic Lorentz currents of definite chirality,
$J^{\dagger}_{\alpha}=\bar{u}\,\mathcal{O}_{\alpha}\,d$ and
$j_{\beta}=\bar{e}\,\mathcal{O}_{\beta}\,\nu$, with
\begin{align}
\mathcal{O}_{V\mp A} &= \gamma^{\mu}(1\mp\gamma_5), \notag\\
\mathcal{O}_{S\mp P} &= (1\mp\gamma_5), \notag\\
\mathcal{O}_{T_{L,R}}  &= \tfrac{i}{2}[\gamma_{\mu},\gamma_{\nu}](1\mp\gamma_5)\,.
\label{eq:ops}
\end{align}

\begin{table*}[t]
\centering
\caption{Correspondence between operator-labelling conventions. The relation
between the Doi--Kotani--Takasugi parameters (``DKT'')~\cite{Doi:1985dx,Doi:1992dm,
Muto1989,Hirsch1994} (defined only for the (axial)vector channels), the
$\upvarepsilon$-basis~\cite{Pas1999,Pas2001,Deppisch2012} and the EFT Wilson
coefficients~\cite{Cirigliano2018,Graf2022}, with the operator dimension in the
low-energy EFT (LEFT) and in SMEFT. The compact symbols in the last column are
used throughout.}
\label{tab:conventions}
\smallskip
\renewcommand{\arraystretch}{1.4}
\begin{tabular*}{\textwidth}{@{\extracolsep{\fill}} l c c c c c c @{}}
\toprule
Operator & LEFT & SMEFT & DKT & $\upvarepsilon$-basis & EFT & This work \\
\midrule
Majorana mass insertion
  & 3 & 5
  & $\chevron{m_{\beta\beta}}/m_e$ & $\chevron{m_{\beta\beta}}/m_e$
  & $m_{\beta\beta}/m_e$ & $\textcolor{cVLL}{\bmu}$ \\
Left-vector
  & 6 & 7
  & $\chevron{\eta}$ & $\VmA$
  & $\tfrac{1}{2}C_{VL}^{(6)}$ & $\textcolor{cVLR}{\beeta}$ \\
Right-vector
  & 6 & 7
  & $\chevron{\lambda}$ & $\VpA$
  & $\tfrac{1}{2}C_{VR}^{(6)}$ & $\textcolor{cVRR}{\blam}$ \\
Scalar
  & 6 & 7
  & --- & $\SP$
  & $\tfrac{1}{2}\bigl(C_{SL}^{(6)}-C_{SR}^{(6)}\bigr)^{\dagger}$
  & $\textcolor{cS}{\bsig}$ \\
Tensor
  & 6 & 7
  & --- & $\TR$
  & $\tfrac{1}{2}C_{T}^{(6)}$ & $\textcolor{cT}{\btau}$ \\
Vector, derivative$^{\dagger}$
  & 7 & 7
  & --- & ---
  & $\tfrac{1}{2}\bigl(C_{VL}^{(7)}-C_{VR}^{(7)}\bigr)$
  & $\textcolor{cS}{\bsig}$ \\
\bottomrule
\end{tabular*}
\\[4pt]
\raggedright\footnotesize
$^{\dagger}$ $\bsig = \tfrac{1}{2}\bigl[C_{SL}^{(6)}-C_{SR}^{(6)}
+\tfrac{m_u+m_d}{v}(C_{VL}^{(7)}-C_{VR}^{(7)})\bigr]$. The derivative operators % $C_{VL,VR}^{(7)}$
enter identically as the scalar operators and are therefore absorbed into their definition. As
$(m_u+m_d)/v\simeq3\times10^{-5}$ makes their contribution numerically small.
\end{table*}

The first term in Eq.~\eqref{eq:Leff} corresponds to the Standard-Model $V\!-\!A$ charged current, the second terms describes new interactions caused by new effective dimension-six operators at the low scale that can be induced from dimension-seven SMEFT operators at a scale $\Lambda\gg v$, with $\upvarepsilon^{\alpha}_{\beta}\propto (v/\Lambda)^{3}$. In the second term of Eq.~\eqref{eq:Leff}, the primed sum runs over all chirality combinations of the leptonic and hadronic currents as specified in Eq.~\eqref{eq:ops} with dimensionless coefficients $\upvarepsilon^{\alpha}_{\beta}$. The $0\nu\beta\beta$ amplitude depends on the time-ordered product of two effective Lagrangians, \cref{eq:ops}, which contains terms linear in $\upvarepsilon^{\alpha}_{\beta}$, 
\begin{equation}
T\left(\mathcal{L}_{(1)} \mathcal{L}_{(2)}\right)\supset\frac{G_F^2}{2} T\left\{\upvarepsilon_\alpha^\beta j_\beta J_\alpha^{\dagger} j_{V-A} J_{V-A}^{\dagger}\right\}\,.
\end{equation}
For double weak nuclear transitions, only five independent combinations survive \cite{Pas1999, Cirigliano2017}, these are the following, where $\upvarepsilon^{\,\mathrm{leptonic}}_{\,\mathrm{hadronic}}$, \\ 
\begin{align}
\Bigl\{\,
\textcolor{cVLL}{\chevron{m_{\beta\beta}}/m_e},\;
\textcolor{cVLR}{\upvarepsilon^{\,V+A}_{\,V-A}},\;
\textcolor{cVRR}{\upvarepsilon^{\,V+A}_{\,V+A}},\;
\textcolor{cS}{\upvarepsilon^{\,S+P}_{\,S\pm P}},\;
\textcolor{cT}{\upvarepsilon^{\,T_R}_{\,T_R}}
\,\Bigr\} \notag\\
\;\equiv\;
\Bigl\{\,
\textcolor{cVLL}{\bmu},\;
\textcolor{cVLR}{\beeta},\;
\textcolor{cVRR}{\blam},\;
\textcolor{cS}{\bsig},\;
\textcolor{cT}{\btau}
\,\Bigr\}\,.
\end{align}

This formalism originated in~\cite{Doi:1985dx,Doi:1992dm}, and the first limits were calculated in~\cite{Muto1989} for $0\nu\mm$. For the positron modes, this was performed first in~\cite{Hirsch1994}. The original works on this subject studied however only the (axial)vector currents $\left\{\bmu, \beeta, \blam\right\}$, and their interferences\footnote{Although derived in the context of the Left-Right Symmetric Model, these resulting rate formulae hold for any theory that generates these effective couplings~\cite{Pati:1974yy, Mohapatra:1974gc}.}.

The modern chiral-EFT \emph{master formula} of~\cite{Cirigliano2017,Cirigliano2018}, however, covers all possible currents, including $\btau$ and $\bsig$, but has been formulated only for $0\nu\mm$, denoted ``EFT'' in \cref{tab:conventions}.

In this work we modify the \emph{master formula} to include $0\nu\pp$ and $0\nu\pec$, and quantify the complementarity of the three neutrinoless modes. We label the \emph{effective couplings} by the chirality of the leptonic current they carry: the (axial)vector effective couplings $\bmu$, $\beeta$, $\blam$, the (pseudo)scalar $\bsig$, and the tensor $\btau$. We refer to Tab.~\ref{tab:conventions} for the mapping between other commonly used formalisms, also employed in this work. Note that $\bmu$ corresponds to the light-neutrino mass exchange. 

\subsection{Rate Formula of $0\nu\beta^{-}\beta^{-}$}
\label{sec:rate}
The inverse half-life for neutrinoless double beta transitions developed  in Refs.~\cite{Cirigliano2017,Cirigliano2018}, keeping only long-range contributions, reads
\begin{align}
\bigl[T^{0\nu}_{1/2}\bigr]^{-1} = g_A^4 \Big\{&
   G_{01}\,|A_\nu|^2 \notag\\
 &+ 4\,G_{02}\,|A_E|^2 \notag\\
 &+ 2\,G_{04}\bigl(|A_{m_e}|^2 + \operatorname{Re}[A_{m_e}^* A_\nu]\bigr) \notag\\
 &+ G_{09}\,|A_M|^2
 \notag\\
&- 2\,G_{03}\operatorname{Re}\bigl[A_\nu A_E^* + 2\,A_{m_e} A_E^*\bigr]\notag\\
&+ G_{06}\operatorname{Re}\bigl[A_\nu A_M^*\bigr]
\Big\}\,.
\label{eq:master}
\end{align}
It consists of the PSFs $G_{0k}$ and the \emph{sub-amplitudes} $A_i$. The PSFs encode the kinematics of the emitted lepton(s) in the Coulomb field of the daughter nucleus and are independent of the new-physics model. The sub-amplitudes $A_i$ collect all contributions (including effective couplings and NMEs) associated with a given outgoing-lepton current\footnote{Here $u\equiv u(p_i)$ denotes the Dirac spinor of an emitted charged lepton.}~\cite{Cirigliano2018, Graf2022}:
$A_\nu\,(\bar{u}P_R C\bar{u}^T)$,
$A_E\,(\bar{u}\gamma^0 C\bar{u}^T)$,
$A_{m_e}\,(\bar{u}C\bar{u}^T)$,
$A_M\,(\bar{u}\gamma^0\gamma_5 C\bar{u}^T)$. The six terms of~\cref{eq:master} correspond to the four squared amplitudes and their two interferences. The remaining constants are the nucleon axial-coupling constant $g_A\simeq1.27$, the vector coupling $g_V=1$, the chiral condensate scale $B_\chi\equiv m_\pi^2/(m_u+m_d)$, the CKM matrix element $V_{ud}$, the nucleon form-factor couplings, $g_M\simeq4.7$, $g_T\simeq0.99$; $g_{T'}$ set to $0$ following the conservative choice of \cite{GKS26} (otherwise undetermined, an $\mathcal{O}(10\%)$ effect on $\btau$), and  the nucleon, electron and pion masses $m_N, m_e, m_\pi$, respectively. 

The NMEs  are overlap integrals of the two-nucleon $\Delta L=2$ transition operators between the parent and daughter nuclear ground states, $0^+\to0^+$.  The NMEs contain the nuclear-structure information and constitute the dominant source of theoretical uncertainty. Each long-range NME is associated with a specific spin-isospin structure of the nucleon transition operator --- Fermi ($M_F$), Gamow--Teller ($M_{GT}$) and tensor ($M_T$) and, in the sub-amplitude decomposition \cite{Cirigliano2017, Cirigliano2018} is further decomposed into axial--axial (AA), axial--pseudoscalar (AP), pseudoscalar--pseudoscalar (PP) and weak-magnetism (MM) form-factor contributions. 
The sub-amplitudes read: 

\begin{widetext}
\begin{align}
A_\nu ={}& \textcolor{cVLL}{\bmu\,\bigl(-V_{ud}^2\bigr)\Bigl[
        -\left(\frac{g_V}{g_A}\right)^2 M_F
        + M_{GT}^{AA}+M_{GT}^{AP}+M_{GT}^{PP}+M_{GT}^{MM}} \notag\\
     &\hphantom{{}\bmu\,\bigl(-V_{ud}^2\bigr)\Bigl[}
        \textcolor{cVLL}{+\, \frac{2m_\pi^2 g_\nu^{NN}}{g_A^2}M_{F,\mathrm{sd}} + M_T^{AP}+M_T^{PP}+M_T^{MM}
        \Bigr]} \notag\\
        &+ \textcolor{cS}{\bsig\,\frac{2V_{ud}\,B_\chi}{m_e}
        \Bigl[\tfrac12 M_{GT}^{AP}+M_{GT}^{PP}+\tfrac12 M_T^{AP}+M_T^{PP}\Bigr]} \notag\\
        &+ \textcolor{cT}{\btau\,\frac{2V_{ud}\,m_N}{m_e}
        \Bigl[\frac{2g_{T'}g_V}{g_A^2}\frac{m_\pi^2}{m_N^2}M_{F,\mathrm{sd}}
        -\frac{8g_T}{g_M}\bigl(M_{GT}^{MM}+M_T^{MM}\bigr)\Bigr]}, \notag\\[2pt]
A_E ={}& -\frac{2V_{ud}}{3}\Bigl\{
        \textcolor{cVLR}{\beeta\bigl[\left(\frac{g_V}{g_A}\right)^2 M_F + \tfrac13(2M_{GT}^{AA}+M_T^{AA})\bigr]}
        + \textcolor{cVRR}{\blam\bigl[\left(\frac{g_V}{g_A}\right)^2 M_F - \tfrac13(2M_{GT}^{AA}+M_T^{AA})\bigr]}\Bigr\}, \notag\\[2pt]
A_{m_e} ={}& \frac{V_{ud}}{3}\Bigl\{
        \textcolor{cVLR}{\beeta\bigl[\left(\frac{g_V}{g_A}\right)^2 M_F - \tfrac13(M_{GT}^{AA}-4M_T^{AA})
        - 3(M_{GT}^{AP}+M_{GT}^{PP}+M_T^{AP}+M_T^{PP})\bigr]} \notag\\
     &\hphantom{\frac{V_{ud}}{3}\Bigl\{}
        + \textcolor{cVRR}{\blam\bigl[\left(\frac{g_V}{g_A}\right)^2 M_F + \tfrac13(M_{GT}^{AA}-4M_T^{AA})
        + 3(M_{GT}^{AP}+M_{GT}^{PP}+M_T^{AP}+M_T^{PP})\bigr]}\Bigr\}, \notag\\[2pt]
A_M ={}& \textcolor{cVLR}{\beeta\,\frac{4g_A}{g_M}\frac{m_N}{m_e}V_{ud}\bigl(M_{GT}^{MM}+M_T^{MM}\bigr)}.
\label{eq:amps}
\end{align}
\end{widetext}

The currents-to-amplitude correspondence, encoded in Eq.~\eqref{eq:amps}, is the key organizing principle for the results presented below. The effective operators $\textcolor{cVLL}{\bmu}$, $\textcolor{cS}{\bsig}$ and $\textcolor{cT}{\btau}$ enter exclusively through $A_\nu$; $\textcolor{cVLR}{\beeta}$ contributes to $A_E$ and $A_{m_e}$ and additionally via weak-magnetism enhancement to $A_M$; whereas  $\textcolor{cVRR}{\blam}$ contributes only to $A_E$ and $A_{m_e}$.
\subsection{Modifications to the Positron Emitting Modes}
\label{sec:crossing}

The three modes depicted in Fig.~\ref{fig:diag-pos} can be described by the same effective Lagrangian in \cref{eq:Leff}. However, they describe distinct isospin transitions: $nn\to pp\,e^-e^-$ versus $pp\to nn\,e^+e^+$ and $e^-pp\to nn\,e^+$, and have different kinematics entering their amplitudes. In the following, we describe how to modify the treatment in~\cref{sec:rate} to reflect this. 

\subsubsection*{Nuclear Matrix Elements}

The positron-mode NMEs~\cite{Hirsch1994, Barea2015} are different to the electron-mode ones~\cite{Horoi2017, Cirigliano2017} (isospin-raising, different daughter nucleus). $0\nu\pp$ and $0\nu\pec$ nonetheless share the same NMEs for a given isotope, differing only through the quantum state of the outgoing leptons, which is parameterized by PSFs \cite{Doi:1992dm, Kotila2013}.

\subsubsection*{Phase-Space Factors}

For $0\nu\pp$, relative to the standard $0\nu\mm$ mode, the \emph{two} outgoing leptons are related by crossing symmetry. Hence, the leptonic $\gamma^0$-bilinear (see \cref{sec:rate}) signs are unchanged and \cref{eq:master} applies. However, the PSFs for $0\nu\pp$ are different because the daughter's Coulomb field repels the two positrons and $4m_e$ of energy is spent creating them, shrinking the phase space.

For $0\nu\pec$, relative to the standard $0\nu\mm$ mode, one of the outgoing electrons is replaced by a captured electron in the initial state. This corresponds effectively to one crossing in $\bar{u}(k_1)\to\bar{v}(k_1)$, flipping the sign of lepton bilinears containing~$\gamma^0$ and hence modifying the interference terms in \cref{eq:master} as~\cite{Doi:1985dx,Doi:1992dm,Hirsch1994}
\begin{align}
&G_{03} \to S_e\,f_e\,G_{03}\,,\\
&G_{04} \to f_e\,G_{04}\,,\\
&G_{06} \to S_e\,G_{06}\,,
\label{eq:crossing}
\end{align}
with $S_e = -1$ for $0\nu\pec$ ($S_e = +1$ for $0\nu\mm$) setting the sign of the $\gamma^0$-bilinear terms. Here,
$f_e = 1 - \tfrac{3\alpha}{2 m_e R_A}$ is the bound-state electron correction (where $\alpha$ is the fine-structure constant and $R_A=1.2\,A^{1/3}$~fm is the nuclear radius)~\cite{Doi:1985dx,Doi:1992dm,Kotila2013}. 
Therefore, while the diagonal terms ($G_{01}$, $G_{02}$, $G_{09}$) are unaffected, the $G_{03}$ and $G_{04}$ of the interference terms receive an $\mathcal{O}(1)$ modification in the $\pec$ mode, strongest for the lightest isotopes.

It is this sign structure, not the PSFs alone, that makes the positron modes qualitatively different probes of multi-current scenarios (\cref{sec:interference}).

% ============================================================================
\subsection{Nuclear Physics Inputs}
\label{sec:inputs}
Throughout, the $0\nu\mm$ results are computed exclusively with the NMEs and PSFs from~\cite{Horoi2017}, in the modern chiral-EFT master formula of~\cite{Cirigliano2017}. The positron modes are evaluated with two independent input sets, a ``\emph{modern}'' one and a ``\emph{legacy}'' one, 
each with independent NMEs and PSFs:

\begin{enumerate}[label=(\roman*)]
\item \textbf{Electron mode $0\nu\mm$:} We consider the four isotopes $\{\,^{76}\mathrm{Ge},\,^{82}\mathrm{Se},\,^{130}\mathrm{Te},\,^{136}\mathrm{Xe}\,\}$, with NMEs and PSFs from~\cite{Horoi2017}, tabulated in, and using the sub-amplitude decomposition of \cref{eq:master,eq:amps} of~\cite{Cirigliano2017} (Tabs.~5 and~3 therein). All five currents are accessible.
\item \textbf{Positron modes $0\nu\pp$ and $0\nu\pec$, ``modern'':} We consider the isotopes
$\{\,^{78}\mathrm{Kr},\,^{96}\mathrm{Ru},\,^{106}\mathrm{Cd},\,^{124}\mathrm{Xe},\,^{130}\mathrm{Ba},\,^{136}\mathrm{Ce}\,\}$, with IBM-2 (interacting boson model) NMEs (isospin-restored)~\cite{Barea2015} and PSFs from~\cite{Kotila2013}, both tabulated in~\cite{GKS26} where the NMEs are calculated with updated single particle energies\footnote{\label{footnote3}We thank Jenni Kotila for providing us with the corrected version of the $^{124}$Xe column of Tab.~3 in~\cite{GKS26}.}.
All five currents are accessible.
\item \textbf{Positron modes $0\nu\pp$ and $0\nu\pec$, ``legacy'':} We consider the same six isotopes, with the pnQRPA (proton--neutron quasiparticle random-phase approximation) NMEs~\cite{Hirsch1994} (Tab.~2 therein) and the PSFs from~\cite{Doi:1992dm} (Tabs.~I and~III). Four currents are accessible: the recoil NME $M_R$ gives an approximation of $\btau$, while $\bsig$ has no counterpart, see \cref{app:other-currents} for more details.
\end{enumerate}
Although the \emph{modern} results should be significantly more accurate, the comparison between the two input sets helps to give an estimate of systematic uncertainties. We find overall a good agreement between both approaches which strengthens the interpretation of our results.
%=============================================================================
\section{Single-current sensitivity}
\label{sec:single}

\begin{figure*}[t]
\centering
\includegraphics[width=1.\textwidth]{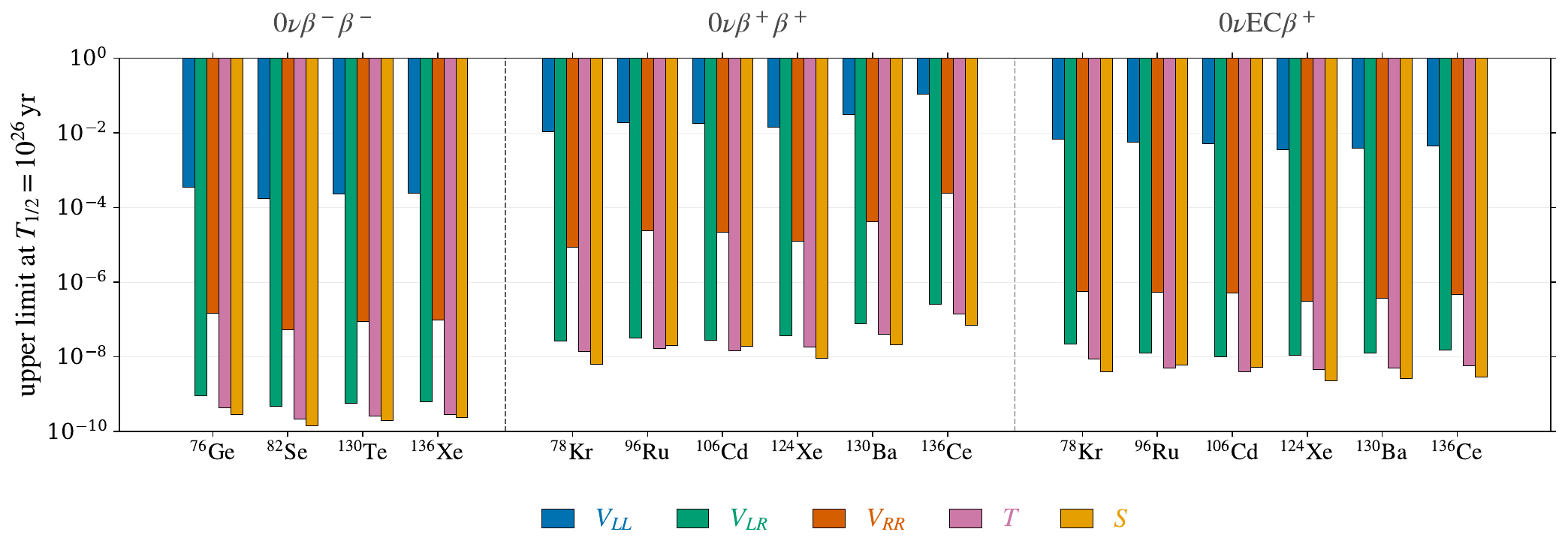}
\caption{Single-current sensitivities at $T_{1/2}=10^{26}$~yr for the electron mode $0\nu\mm$ (left block), and both positron modes, $0\nu\pp$ (middle block) and $0\nu\pec$ (right block), with the modern nuclear inputs.}
\label{fig:sens-new}
\end{figure*}

In the following, we first present the intrinsic ``single-parameter'' limits: the upper limit on each current under the assumption that it is the only one that is non-zero, meaning the $0\nu$ mode is triggered entirely by a single low-energy operator (called ``on-axis'' in the original works~\cite{Muto1989,Hirsch1994}).
In \cref{fig:sens-new}, we use only the modern nuclear inputs and show the results for all three decay modes at a value of $T_{1/2}=10^{26}$~yr, a common reference near the current experimental reach of $0\nu\mm$, isolating the pure NME and PSF dependence. Two patterns are immediately visible:
\paragraph{Within each decay mode, the five currents separate into two distinct tiers,} with bounds differing by two to three orders of magnitude. The two weaker currents, $\cmu$ and $\clam$, are constrained much less stringently than the three stronger ones, $\ceta$, $\ctau$ and $\csig$. This hierarchy originates from explicit enhancement in the latter: The effective couplings $\ceta$ and $\ctau$ are enhanced by the nucleon-to-electron mass ratio $m_N/m_e$ (only one of the $\ceta$ sub-amplitudes carries this enhancement; its remaining pieces have no large factors such that the effect is sub-amplitude specific), and $\csig$ carries the chiral factor $B_\chi/m_e = m_\pi^2/[(m_u+m_d)\,m_e]$. As each of these factors is $\gg 1$, even a small coupling leads to a sizeable rate, and hence the half-life constraint translates into a tight bound. By contrast, the two weaker currents are not enhanced with any large factors, and therefore lead to substantially weaker bounds.

\paragraph{The positron modes are intrinsically less sensitive probes than $0\nu\mm$ decay,} since producing one or two positrons costs rest-mass energy, reducing the available $Q$ value and thereby shrinking the phase space. This reduced sensitivity, however, is far from uniform across the different currents. Comparing $0\nu\pec$ ($^{124}$Xe)  with $0\nu\mm$ ($^{136}$Xe), the intrinsic sensitivity weakens by a factor $\sim\mathcal{O}(10)$ for $\cmu$, $\ceta$, $\ctau$ and $\csig$, but by only a factor $\sim3$ for $\clam$. The reason for this difference is kinematic. The bound-state electron captured in $0\nu\pec$ modifies the relative weights with which the individual phase-space factors $G_{0k}$ enter the rate (cp. \cref{eq:crossing}), enhancing the combinations $G_{02}$ and $G_{09}$ that multiply $\clam$, while suppressing most strongly $G_{01}$, the factor multiplying the mass mechanism, $\cmu$. Consequently, $0\nu\pec$ does not simply test the same underlying physics with reduced sensitivity---it provides a \emph{selective} probe of the right-handed current $\clam$. As $0\nu\pp$ does not involve any captured electrons, it results in no PSF reweighting, such that the phenomenology is much closer to $0\nu\mm$, however, with a lower sensitivity.

%=============================================================================
\section{Target sensitivity of $0\nu\pec$}
\label{sec:nudoubt}
%=============================================================================

 At equal half-life the positron modes always lag behind $0\nu\mm$ (\cref{fig:sens-new}). However, under the assumption of the same underlying effective coupling, in case of an observation of $0\nu\mm$, a half-life can be predicted at which a $0\nu\pec$ observation is expected (at which the sensitivity of $0\nu\pec$ becomes ``competitive''). As an example, we identify the corresponding half-life at which $0\nu\pec$ probes the same coupling strength as the current best $0\nu\mm$ limit, assuming one effective coupling $c_i$ at a time. This defines a channel-by-channel benchmark for the positron programme.

Under these assumptions, the master formula in \cref{eq:master} reduces to a simple proportionality between the decay rate of a given process $X$ and the square of the corresponding effective coupling,
\begin{align}
\frac{1}{T^{0\nu}_{1/2}[X]} &\;=\; C_i[X]\,
c_i^2 \,,\label{eq:single}\\
\qquad c_i &\in \{\,\bmu,\ \beeta,\ \blam,\ \btau,\ \bsig\,\}\nonumber,
\end{align}
where $X$ denotes the decay mode for a specific isotope, and $C_i[X]$ is the diagonal of the rate matrix for a specific effective coupling. With a single coupling at a time, no interference terms can arise such that only the diagonal element enters, which comprises all relevant NMEs and PSFs. 

\begin{figure}[t]
\centering
\includegraphics[width=\columnwidth]{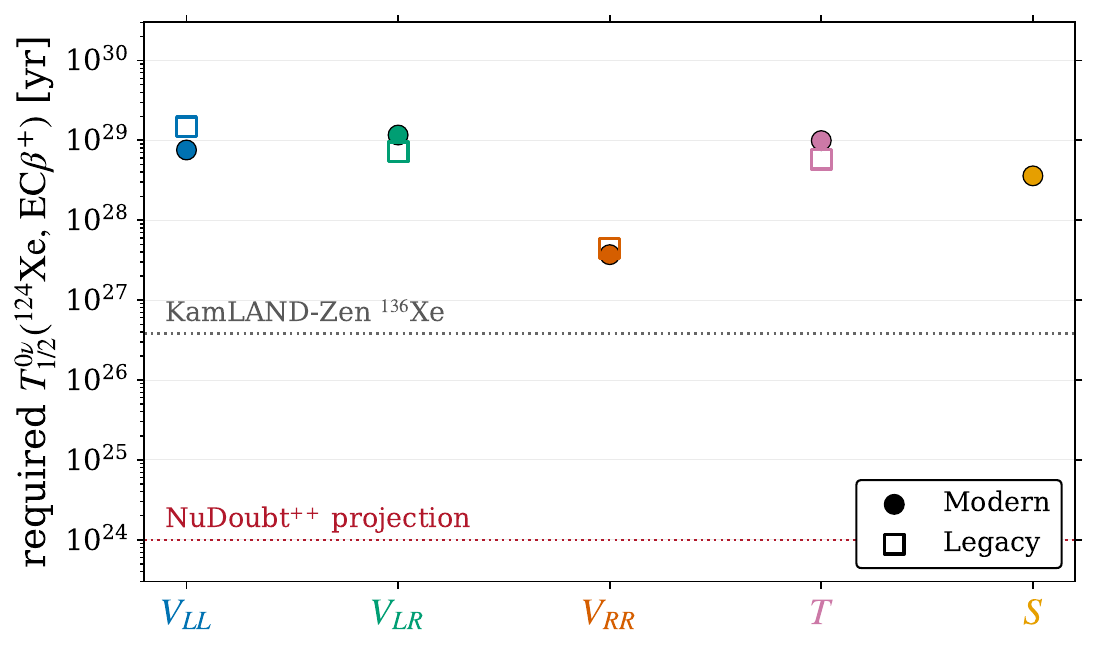}
\caption{Half-life sensitivity required on $^{124}$Xe $\pec$ to match the KamLAND-Zen 2024 limit on $^{136}$Xe (using through the master formula~\cite{Cirigliano2017,Cirigliano2018}) in each current channel, for the modern (filled circles) and legacy (open squares) positron-mode data inputs. The grey dotted line marks the current KamLAND-Zen $^{136}$Xe half-life~\cite{KamLAND-Zen:2024eml}; the red dotted line marks the NuDoubt$^{++}$ projected sensitivity of $10^{24}$~yr~\cite{NuDoubt:2024jax}.}
\label{fig:124Xe-crossover}
\end{figure}

Hence, a certain half-life reach $T^{0\nu}_{1/2}$ excludes couplings above $c_i = 1/\sqrt{C_i[X]\,T^{0\nu}_{1/2}}$\,, which we refer to as the \emph{intrinsic sensitivity} of $X$ to the
current $i$.
In order identify which half life of $0\nu\pec$ in $^{124}\mathrm{Xe}$ is required to achieve the same sensitivity in effective couplings as $0\nu\mm$ in $^{136}\mathrm{Xe}$, we identify from Eq.~\eqref{eq:single} the
\emph{crossover half-life} by demanding
\begin{equation}
T^{\pec}_{124} =
T^{\mm}_{136}\,
\frac{C^{\mm}_{i,136}}{C^{\pec}_{i,124}}\,,
\label{eq:crossover}
\end{equation}
with the shorthand notation $T^{0\nu}_{1/2}[^{124}\mathrm{Xe}, \pec] = T^{\pec}_{124}$.
The most stringent current half-life limit on any $0\nu\beta\beta$ isotope is currently
$T^{\mm}_{136} > 3.8\times 10^{26}\,\mathrm{yr}$ from KamLAND-Zen~\cite{KamLAND-Zen:2024eml}. Translating this half-life into coupling bounds via \cref{eq:single}, with the electron-mode inputs of \cref{sec:single}, yields
\begin{equation}
\begin{aligned}
\bmu   &\lesssim 6.5\times 10^{-8}, \\
\beeta &\lesssim 6.5\times 10^{-10}, \qquad&
\blam  &\lesssim 1.0\times 10^{-7}, \\
\btau  &\lesssim 2.9\times 10^{-10}, &
\bsig  &\lesssim 2.4\times 10^{-10}\,.
\end{aligned}
\label{eq:KZ-best}
\end{equation}
Similarly obtained are the $^{136}$Xe entries of Fig.~\ref{fig:sens-new} (assuming $10^{26}$~yr) by rescaling $(10^{26}/3.8\times10^{26})^{1/2}\simeq0.5$ from to the KamLAND-Zen limit.
These set the targets against which the proposed $\pec$ programme of NuDoubt$^{++}$~\cite{NuDoubt:2024jax} must be compared: In \cref{fig:124Xe-crossover} we show the necessary crossover half-life necessary in $0\nu\pec$ to achieve the same sensitivity in effective couplings, for both the modern (filled circles) and legacy (open squares) positron inputs. Interestingly, the right-handed current $\blam$ is the one that is obtained the earliest at
$T^{\pec}_{124} \sim 4\times 10^{27}\,\mathrm{yr}$, only a factor $\sim 10$ above the current KamLAND-Zen $^{136}$Xe sensitivity. This factor is remarkably stable against the choice of nuclear inputs: it shifts by only $\sim20\%$ between the modern and legacy nuclear input, whereas e.g.\ the $\cmu$ crossover moves by a factor of a few through the IBM-2 vs pnQRPA mass NME. By contrast $\bsig$ 
% \jh{do you want to maintin the color for the currents throughout the paper? It stops here}
requires $\sim 4\times 10^{28}\,\mathrm{yr}$, and $\bmu$, $\beeta$ and $\btau$ all $\sim 10^{29}\,\mathrm{yr}$, orders of magnitude beyond the current $0\nu\mm$ sensitivity. Similar is obtained for the isotope $^{78}$Kr: A crossover sensitivity of $\sim 1.2$--$1.7\times 10^{28}$~yr is obtained for $\blam$ and $\gtrsim 10^{29}$~yr for the other channels. The hierarchy is a property of the decay mode, not of the particular isotope, with $^{124}$Xe being the most favourable, owing to its comparatively large $Q$-value and nuclear charge, which maximise the $\pec$ phase space among the six candidates.

We stress that this does not make NuDoubt$^{++}$ a competitor in raw reach: by the time the $0\nu\pec$ programme approaches the $10^{27}$~yr regime, the $0\nu\mm$ experiments will have moved beyond it, raising the bar further. Its role is that of a \emph{diagnostic}. If a $0\nu\mm$ signal is observed, the crossover half-life quantifies exactly when a $0\nu\pec$ observation is expected given a certain current contributes. Given the large hierarchies between the right-handed current $\blam$ and the others, $0\nu\pec$ can be used to probe the latter.
If instead no $0\nu\mm$ is observed and the $0\nu\mm$ limits keep improving, the required $0\nu\pec$ half-life grows with them. Compared to~\cite{Hirsch1994}, where the comparative strength of the positron modes for $\blam$ was already visible, we present here the quantitative crossover benchmark, its evaluation for the NuDoubt$^{++}$ isotopes, and the inclusion of all five currents.

%=============================================================================
\section{Ratios with positron modes as a discrimnator}
\label{sec:DPratios}
%=============================================================================
\subsubsection*{Single Current Analysis}

\begin{figure*}[t]
\centering
\includegraphics[width=\textwidth]{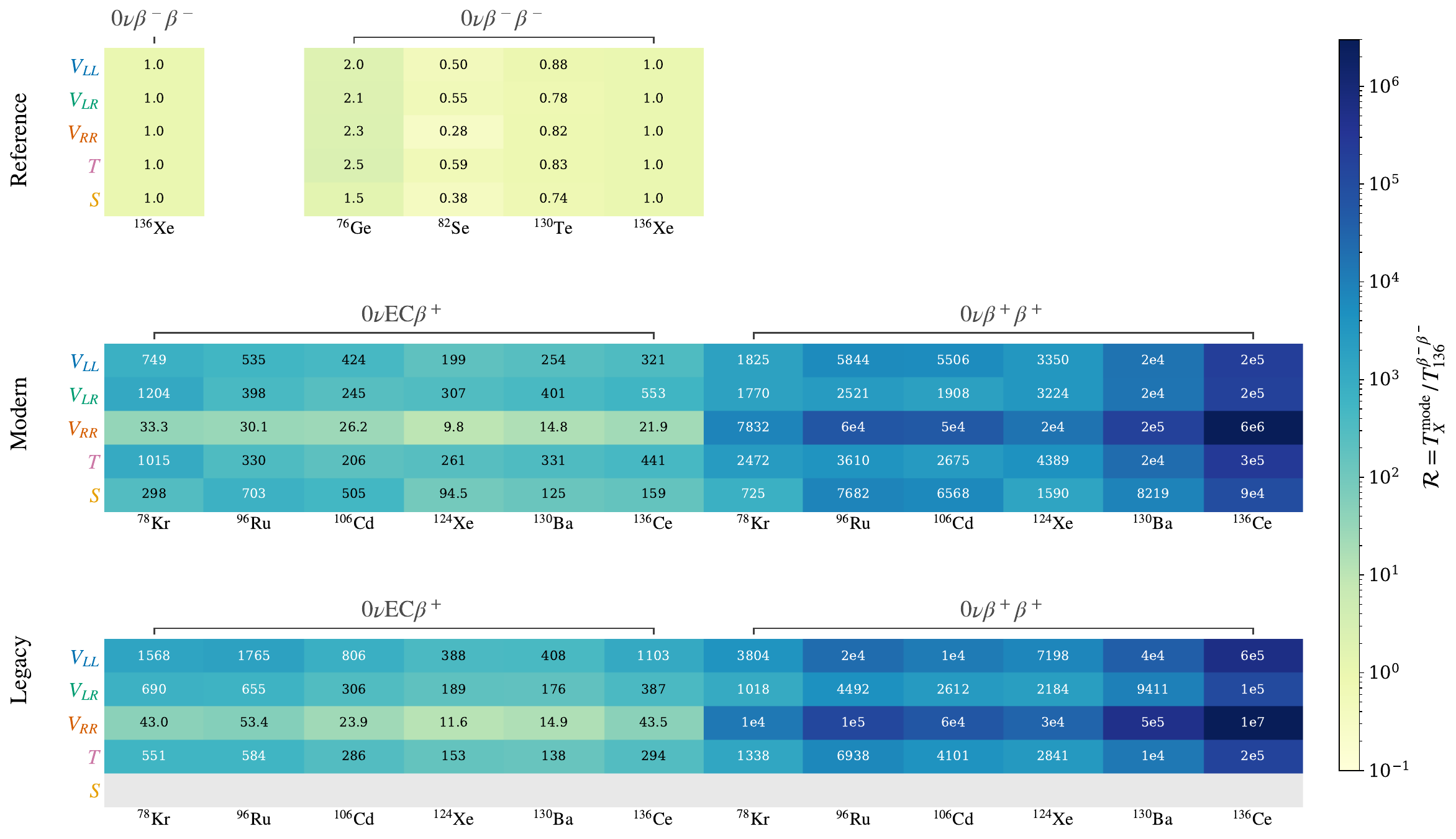}
\caption{Single-current half-life ratios
$\mathcal{R}_i = T_{1/2}[X]/T_{1/2}[^{76}\mathrm{Ge},\mm]$, \cref{eq:Rc}, for every current (rows) and measurement $X$ (columns). \textbf{Top:} $\mm$ isotopes (NMEs and PSFs of~\cite{Horoi2017}, tabulated in ~\cite{Cirigliano2017}), with the $^{76}$Ge reference strip at unity. \textbf{Middle:}  $\pec$ and $\pp$ with the modern data version (~\cite{GKS26}: IBM-2 NMEs with isospin restoration~\cite{Barea2015}, PSFs following ~\cite{Kotila2013}). \textbf{Bottom:}  the same with the legacy data version (NMEs~\cite{Hirsch1994}, PSFs~\cite{Doi:1992dm}), which does not give access to $\bsig$ (grey band).}
\label{fig:Rc-atlas}
\end{figure*}

In the following, we want to entertain the possibility that in the future \emph{two} half-lives may have been measured---or one measured and one constrained---and we investigate what this information is able to reveal about the underlying current. Taking half-life \emph{ratios} can be useful as the NMEs and PSFs differ for different effective operators. Their corresponding ratios take different values whilst the unknown effective coupling $|c_i|^2$ \emph{cancels} out. This was first proposed in~\cite{Deppisch:2006hb} as a new method to discriminate LNV operators in $0\nu\mm$ decay with the help of different isotopes. Furthermore, the work in \cite{Graf2022} evaluates isotope ratios and kinematic observables to distinguish between 32 low-energy effective operators, and~\cite{Agostini:2022bjh} shows that it can sometimes be necessary to consider a third isotope to break degeneracies between the mass mechanism and other exotic mechanisms.

In this section, we show that considering ratios of half-lives of \emph{different modes} is similarly powerful.
The ratio $\mathcal{R}_i$ of half-lives assuming a single effective coupling $c_i$ dominant at a time,
\begin{align}
    \mathcal{R}_i=\frac{C^{\mm}_{i,136}}{C^{\pec}_{i,124}}=\frac{T^{\pec}_{124}}{T^{\mm}_{136}}\,,
\label{eq:Rdef}
\end{align}
is a pure ratio of the coefficients $C_i$, evaluated with the electron-mode inputs for $^{136}$Xe and the positron inputs for $^{124}$Xe. 
However, taking ratios $\mathcal{R}_i$ of $0\nu\mm$ half-lives of different isotopes, as proposed in~\cite{Deppisch:2006hb}, turns out to separate the currents only weakly. This is visualized in Fig.~\ref{fig:Rc-atlas} (upper block), where we compare the \emph{predictions} of $\mathcal{R}$ for different currents and isotopes for $0\nu\mm$. In order to discriminate among effective couplings, $\mathcal{R}_i$ has to differ appreciably from current to current. However, for $0\nu\mm$ isotope pairs this is only moderately achieved. Across all isotopes the $\beeta$ prediction agrees with that of $\cmu$ to within $\lesssim30\%$, and even $\blam$ stays within a factor $\lesssim2$ for most isotopes. The order-of-magnitude discrimination that was found for $\blam$ \cite{Deppisch:2006hb} relies in particular on a single pair: the strongly-deformed $^{150}$Nd, whose NME should be taken with caution \cite{Deppisch:2006hb}, and $^{128}$Te, whose small $Q$-value makes it experimentally challenging.

This is in contrast when we consider ratios of two different modes, i.e. the above defined ratio $\mathcal{R}$. As the bound electron reshapes the phase space of $0\nu\pec$ with respect to the ones of $0\nu\mm$, the predictions for $\mathcal{R}$ spread over a wider range. Using the modern positron inputs, the predictions for the five different currents using $^{124}$Xe\footnote{With the legacy inputs we obtain $\mathcal{R}_{\blam}\sim12$, $\mathcal{R}_{\btau}\sim150$, $\mathcal{R}_{\beeta}\sim190$, $\mathcal{R}_{\bmu}\sim390$, with no bound on $\bsig$.}
\begin{align}
  % \mathcal{R}_{\blam} &\sim 11\,,\\
  % \mathcal{R}_{\bsig}&\sim  107\,,\\
  % \mathcal{R}_{\bmu} &\sim 219\,,\\
  % \mathcal{R}_{\btau} &\sim  293\,,\\
  % \mathcal{R}_{\beeta} &\sim 344\,.
    \mathcal{R}_{\blam} &\sim 10\,,\notag\\
  \mathcal{R}_{\bsig}&\sim  95\,,\notag\\
  \mathcal{R}_{\bmu} &\sim 199\,,\notag\\
  \mathcal{R}_{\btau} &\sim  261\,,\notag\\
  \mathcal{R}_{\beeta} &\sim 307\,.
\label{eq:DPratios}
\end{align}
Most interestingly, $\mathcal{R}_{\blam}$ lies nearly an order of magnitude below all the others (in both datasets).

Figure~\ref{fig:Rc-atlas} applies this approach to all sixteen possible combinations: the four electron-mode isotopes and the six positron-mode isotopes in both decay channels, each using both positron datasets. Each cell shows the single-current half-life ratio with $^{136}$Xe:
\begin{equation}
\mathcal{R}_i[X] \;=\; \frac{T_X^{\text{mode}}}{T^{\mm}_{136}}
\;=\; \frac{C^{\mm}_{i,136}}{C^{\text{mode}}_{i,X}}\,,
\label{eq:Rc}
\end{equation}
i.e.\ the half-life at which for a certain isotope and effective coupling $c_i$ (one dominating at a time) an observation would be expected given a discovery of $0\nu\mm$ in $^{136}$Xe at a certain half-life. The discriminating power of a possible pair of measurements is depicted as color difference in each column: if all five entries were equal, the ratio would carry no information on the mechanism at all. 
In the $0\nu\mm$ block (top) the columns are nearly uniform: exchanging one $\mm$ isotope for another changes $\mathcal{R}_i$ by factors of order unity---the original proposal in~\cite{Deppisch:2006hb} within a single mode has limited impact. 

As emphasized in~\cite{Deppisch:2006hb}, within a single mode, common NME uncertainties cancel in the ratio. Across modes this cancellation does not occur to the same degree, since mother and daughter nuclei and the isospin structure of the transition differ\footnote{\label{footnote5}Note that the three sets of nuclear inputs we use in this work are all calculated with different techniques (pnQRPA, shell-model and IBM-2), so we expect the theoretical uncertainties to not cancel \emph{a priori}. We comment on how the results change when one changes the nuclear inputs in~\cref{app:depend}. 
% We believe it is useful to employ a variety of possible nuclear inputs.
}. 
In the $0\nu\pec$ and $0\nu\pp$ blocks (middle) the larger change in tones of colour indicates a wider spread in expected ratios $\mathcal{R}$: in $0\nu\pec$ the right-handed current $\blam$ is one to two orders of magnitude apart from the rest. As there is an intrinsic uncertainty to the NME calculations, we try to give an estimate on the residual nuclear-structure systematics by presenting results for two independent nuclear inputs. Hence the bottom block shows the same analysis using the legacy inputs (NMEs~\cite{Hirsch1994},
PSFs~\cite{Doi:1992dm}; no $\bsig$): the observation remains unchanged, with legacy and modern differing only by $\mathcal{O}(1)$ factors such that the discrimination seems to be robust against changes in the nuclear-structure input.

This highlights the complementary information obtained by $0\nu\pec$. Suppose $0\nu\beta\beta$ is discovered in $^{76}$Ge at $T_{1/2}=10^{28}$~yr. According to the $^{124}$Xe $\pec$ column of \cref{fig:Rc-atlas}, the mass mechanism, $\cmu$, predicts a $0\nu\pec$ signal only at $\sim1\times10^{30}$~yr, whereas a purely right-handed current, $\clam$, predicts one at $\sim5\times10^{28}$~yr---twenty times earlier. While a possible observation is clearly expected first in $0\nu\mm$, an $0\nu\pec$ observation at a few $10^{28}$~yr would be a smoking gun signal for a right-handed current, while a null result at that level would exclude $\clam$ dominance of the $^{76}$Ge signal. An analogous statement holds for every column of \cref{fig:Rc-atlas}.

\subsubsection*{Multiple Current Analysis}
\label{sec:no-interference}

So far, we considered a single current at a time. However, the observed neutrinoless weak decay signal can have contributions from all the possible currents. However, the large spread between the single-current ratios in~\cref{sec:single} can be useful, even with multiple currents contributing. This is clearest at the extremes, where for our benchmark scenario, an observed $\mathcal{R}_i$ over 261 would indicate the presence of $\beeta$. However, this is the longest lifetime and therefore, the most experimental challenging target. On the other extreme, where an observed $\mathcal{R}_i<95$ would establish a contribution from $\blam$ without knowing the value of the effective coupling, in the most experimental accessible regime.

In the following, we assume an eventual $0\nu\mm$ \emph{observation}, such that the positron modes act again as a \emph{discriminator} of the underlying current, providing information that no $0\nu\mm$ isotope alone can supply.

We will show that even a \emph{positron-mode non-observation} constrains the current present in the decay, especially the right-handed current, $\blam$.

We allow for several currents to be present at the same time, while neglecting interference, and will leave the full treatment of interferences for \cref{sec:interference}, where we will find that this approximation is justified in certain cases. In that case, the $0\nu\pec$ half-life can be written as a sum of individual contributions of the different currents similar to Eq.~\eqref{eq:crossover},
\begin{equation}
\frac{1}{T^{\pec}} \;=\; \frac{1}{T^{\mm}} \sum_i \frac{f_i}{\mathcal{R}_i}\,,
\label{eq:mixrate}
\end{equation}
where $f_i \equiv T^{\beta\beta} C_i c_i^2$ is the fraction of the observed $0\nu\mm$ carried by a specific effective coupling $c_i$, where $\displaystyle\sum_i f_i = 1$\footnote{A given ultraviolet model that generates $c_i$ may in addition induce a Majorana mass through loop mixing into the Weinberg operator~\cite{GrafHati2025}; such a contamination simply appears as a $\bmu$ component of the admixture and is automatically included.}.

Therefore, in the case of a $0\nu\mm$ observation, and a (non-)observation of $0\nu\pec$ at $T^{\pec}\ge T_{\rm lim}$ this sets a limit of $1/r \equiv T^{\mm}/T_{\rm lim}$,

\begin{equation}
\sum_i \frac{f_i}{\mathcal{R}_i} \;\le\; \frac{1}{r}\,.
\label{eq:mixlimit}
\end{equation}

For example, for a fixed right-handed share $f_{\blam}$, the remaining weight $1-f_{\blam}$ may be distributed arbitrarily among the other four currents. The resulting $0\nu\pec$ rate is smallest when that share is placed entirely on the current with the largest ratio, in particular $1/\mathcal{R}_i \ge 1/\mathcal{R}_{\beeta}$ for every $i$. Therefore, the most conservative limit on $f_{\blam}$, its least constraining configuration, would be
$f_{\beeta}=1-f_{\blam}$ with all remaining $f_i=0$. This 
yields the upper bound of
\begin{equation}
\begin{aligned}[b]
&\frac{f_{\blam}}{\mathcal{R}_{\blam}}
 + \frac{1-f_{\blam}}{\mathcal{R}_{\beeta}}
 \le \frac{1}{r}\\
&f_{\blam}
 \le
 \frac{1/r - 1/\mathcal{R}_{\beeta}}
      {1/\mathcal{R}_{\blam} - 1/\mathcal{R}_{\beeta}}
 \equiv f_{\blam}^{\mathrm{lim}}\,,\\
&\Longrightarrow\quad
c_{\blam}^2
\le
\frac{C_{\blam}}{T^{\mm}}f_{\blam}^{\mathrm{lim}}\,.
\end{aligned}
\label{eq:flam}
\end{equation}
This bound is valid for an admixture of any number of
currents. As it is saturated only by the two-current combination $(\blam,\beeta)$, every genuine multi-current admixture yields a strictly
stronger constraint on $f_{\blam}$ and hence on $c_{\blam}$. The only assumption entering this bound is the neglect of interference, which we examine in~\cref{sec:interference}.

Therefore, even a non-observation of the positron mode would begin to give valuable information on the composition of a $0\nu\mm$ signal, which cannot be easily achieved with $0\nu\mm$ isotope ratios only.

%=============================================================================
\section{Degeneracies and discriminating power across modes}
\label{sec:interference}
%=============================================================================
\begin{figure*}[t]
\centering
\includegraphics[width=\textwidth]{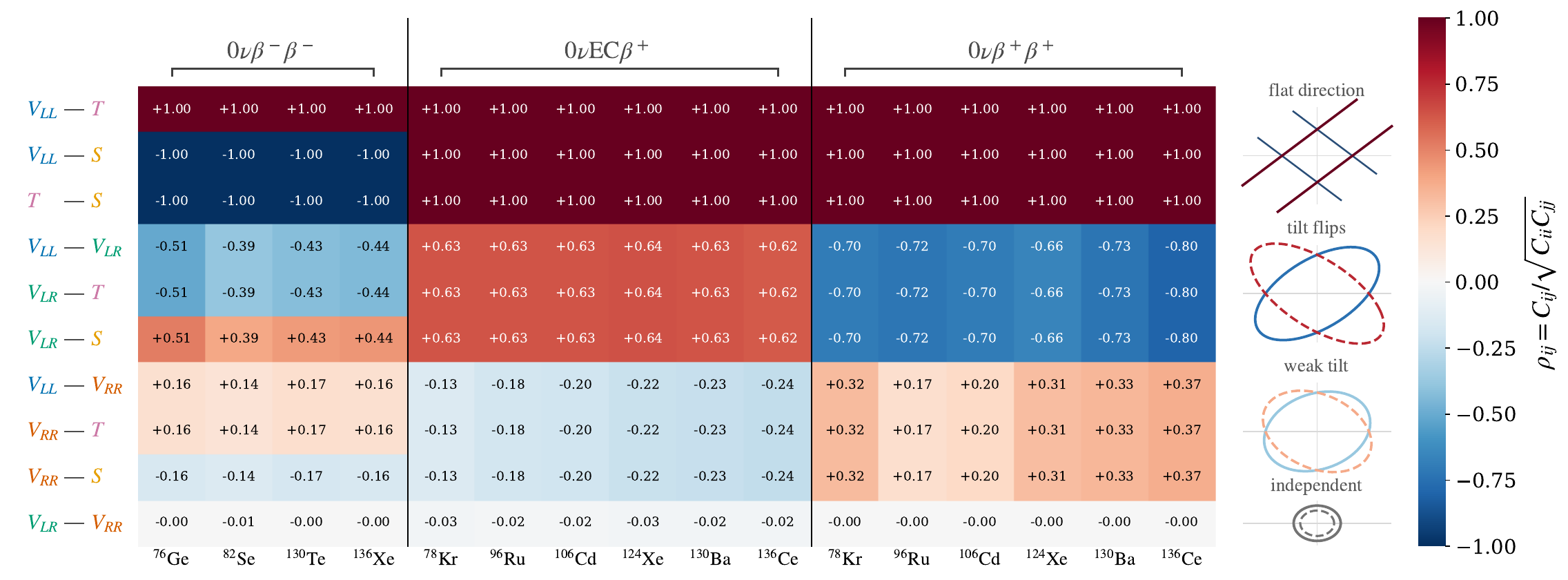}
\caption{Matrix of pairwise interference. Each cell shows the normalized interference coefficient $\rho_{ij}$, \cref{eq:rho}, for each pair of currents (rows) and each isotope/mode combination (columns). The schematic pictures on the right show the corresponding constraint geometry for each family of results in a two-current plane.}
\label{fig:rho-atlas}
\end{figure*}

Once several currents are at play and interferences between different currents are considered, a single measured half-life defines the surface of an ellipsoid in the space of the non-zero effective couplings.

A general quadratic form for this ellipsoid reads\footnote{We drop the labels and write $C_{X}^{\pec} \equiv C$, $T_{X}^{\pec} \equiv T$.}
\begin{equation}
\frac{1}{T} \,=\, \sum_{ij} C_{ij}\,c_i c_j\,,
\quad c = (\bmu,\beeta,\blam,\btau,\bsig)\,,
\label{eq:ratematrix}
\end{equation}
with a half-life measurement being completely characterized by the symmetric matrix $C$. Its diagonal reproduces the single-current sensitivities of \cref{sec:single}, while each off-diagonal entry $C_{ij}$ measures the interference between two currents. The size of $C_{ij}$ depends on the arbitrary normalization of each coupling (the conventions of \cref{tab:conventions}): rescaling a single coupling, $c_i\to a\,c_i$ with the others held fixed, leaves the rate invariant and hence sends $C_{ii}\to C_{ii}/a^2$ and $C_{ij}\to C_{ij}/a$ for $j\neq i$. Therefore, dividing by the geometric mean of the two diagonals cancels this rescaling, leaving a dimensionless, convention-independent \emph{pairwise interference} coefficient
\begin{equation}
\rho_{ij} \;=\; \frac{C_{ij}}{\sqrt{C_{ii}C_{jj}}} \;\in\; [-1,1]\,,
\label{eq:rho}
\end{equation}
whose range is bounded to $[-1,1]$ by the positive-definiteness of $C$. See also \cite{Graf2022} for a similar approach.

In the rest of this section, we study the case of only two effective couplings contributing at once (setting the remaining three to zero), where the constraint reduces to $C_{ii}c_i^2 + 2C_{ij}c_ic_j + C_{jj}c_j^2 = 1/T$ and $\rho_{ij}$ fixes the shape of the allowed ellipse. In \cref{fig:rho-atlas}, we present these pairwise combinations for all ten current pairs and sixteen isotope/mode combinations.

If $\rho_{ij}=0$ the two currents add without an interference term, so the limit on either coupling is unaffected by the presence of the other, as assumed in \cref{sec:no-interference}. For $|\rho_{ij}|\to1$ an exact flat direction exists along which the rate vanishes and no limit can be obtained, whereas an intermediate $\rho_{ij}$ signals a partial correlation.
The sign of $\rho_{ij}$ in \cref{fig:rho-atlas}, encoded in the colour of each cell, fixes the tilt of the allowed ellipse in the $(c_i,c_j)$ plane. A
\textcolor{rhopos}{red} cell ($\rho_{ij}>0$) marks a pair that interferes destructively---the couplings have opposite signs, and the ellipse stretches in the direction of $c_j=-c_i$. A
\textcolor{rhoneg}{blue} cell ($\rho_{ij}<0$) is the mirror image, the currents interfere constructively and the ellipse is along $c_j=c_i$.

\cref{fig:rho-atlas} exhibits four geometries of current pairs, into which its rows are sorted from top to bottom (with the schematic ellipses on the right), here (a)-(d):\\

\emph{(a) Exact degeneracies:} the pairs $\bmu$--$\btau$, $\bmu$--$\bsig$ and $\btau$--$\bsig$ feature $|\rho_{ij}|=1$ in every column. Recall that the rate matrix $C$ is built from the sub-amplitudes of \cref{eq:amps}. The effective couplings $\bmu$, $\btau$ and $\bsig$ enter \emph{only} through $A_\nu$ as a linear combination of the three couplings. Therefore, any variation of $(\bmu,\btau,\bsig)$ that holds $A_\nu$ unaltered, also leaves the rate unchanged. This means a single measurement determines only the linear combination of
the three couplings; for two fixed couplings the third one can always be chosen to obtain a certain $A_\nu$. Breaking this degeneracy requires several isotopes with significantly different coefficient ratios in $A_\nu$ or, more efficiently, a different decay mode.\\

\emph{(b) Mode-dependent interference:} the $\beeta$ rows
($\bmu$--$\beeta$, $\beeta$--$\btau$, $\beeta$--$\bsig$) are sizable and flip sign depending on the decay mode. The flip between $0\nu\mm$ and $0\nu\pec$ is the direct imprint of the crossing of signs in \cref{eq:crossing}: $S_e=-1$ reverses the $G_{03}$ and $G_{06}$ cross-terms through which $\beeta$ interferes with the currents contained in $A_\nu$. Consequently, the degenerate direction in the $(\bmu,\beeta)$ plane \emph{rotates} when the decay mode changes, unlike any combination of $\mm$ isotopes.\\

\emph{(c) Weak interference:} the $\blam$--$A_\nu$ rows ($\bmu$--$\blam$, $\blam$--$\btau$, $\blam$--$\bsig$) are uniformly small.\\

\emph{(d) Decoupling:} $\rho_{\beeta\blam}\approx 0$ everywhere as $\beeta$ and $\blam$ can be treated as effectivly independent, shown by the near-vanishing $\rho_{\beeta\blam}$ in~\cref{fig:rho-atlas}. The reason behind is that in the sub-amplitudes of~\cref{eq:amps}, $\blam$ contributes to $A_E$ and $A_{m_e}$, but not to the weak-magnetism amplitude
$A_M$, where only $\beeta$ contributes. The latter, however, carries the enhancement factor $(m_N/m_e)^2g_M^2$ and saturates $C_{\beeta\beeta}$ alone. Within the sub-amplitudes the two currents do share they are
almost exactly anticorrelated: setting $A_M\to0$ gives
$\rho_{\beeta\blam}=-0.99$ for $^{136}${Xe} and $-1.00$ for
$^{124}${Xe}. As $A_M$ dominates but contains no dependence on $\blam$, it is therefore a good approximation to treat the two currents as independent, as their overlap is numerically negligible.

Families (c) and (d) together make the right-handed current an interference-robust target and hence confirm the assumptions made in~\cref{sec:single}. The current $\blam$ interferes only weakly with the $A_\nu$ currents (c) and negligibly with $\beeta$ (d). However, since 
$\mathcal{R}$ is no longer a weighted average of
the single-current values, the argument of $\mathcal{R}<95$ ensuring the presence of $\blam$ no longer holds. In the following, we see to what extent this threshold is affected.

\begin{figure}[t]
\hspace*{-1cm}\includegraphics[width=0.56\textwidth]{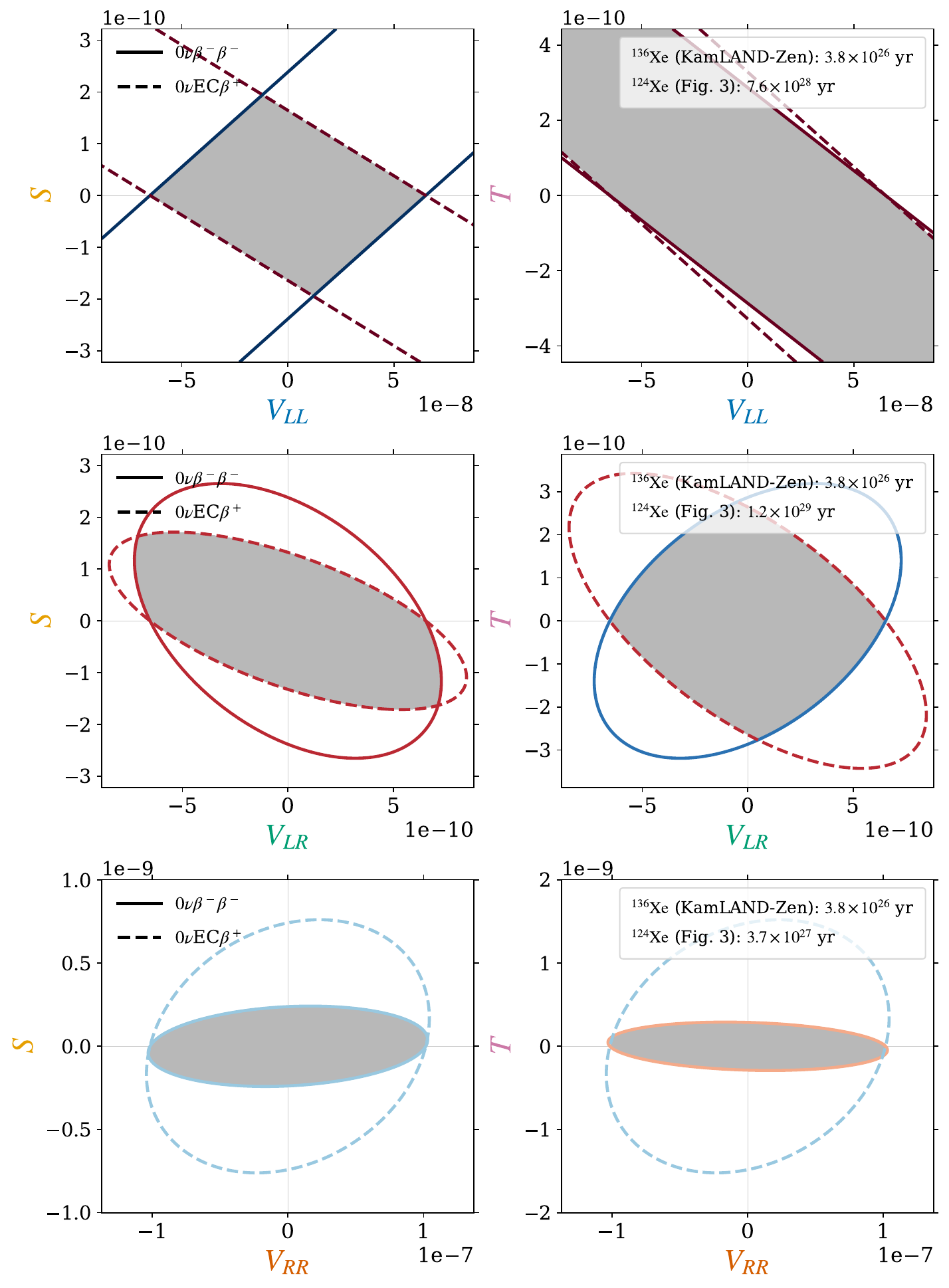}
\caption{Joint constraints in the scalar ($\bsig$) and tensor ($\btau$) planes for the
three (axial-)vector currents $\bmu,\beeta,\blam$. The $^{136}$Xe $\mm$ limit is fixed at the KamLAND--Zen value \cite{KamLAND-Zen:2024eml} ($3.8\times10^{26}$~yr, \emph{solid}); 
the $^{124}$Xe $\pec$ contour (\emph{dashed}) is drawn at each current's crossover half-life
(\cref{fig:124Xe-crossover}). The grey region is allowed by both limits. The contour colours matches \cref{fig:rho-atlas}.}
\label{fig:planet}
\end{figure}

For illustration, we present some of these geometries in~\cref{fig:planet}, using the values from~\cref{fig:124Xe-crossover} for demonstration. This type of 2D current plot, which highlights interference, was originally first presented in~\cite{Hirsch1994} for the three (axial-)vector currents ($\bmu$, $\beeta$ and $\blam$). Here we confront those currents with other possibilities, namely the scalar $\bsig$ and tensor $\btau$ currents. In every panel of~\cref{fig:planet} the \emph{solid} contour depicts the $^{136}$Xe $\mm$  KamLAND--Zen limit $T_{1/2}=3.8\times10^{26}$~yr. The dashed contours are the $^{124}$Xe $\pec$ constraints drawn at the crossover half-lives, which  correspond to the $\pec$ sensitivity at which $0\nu\pec$ would equal the $0\nu\mm$ bound for each current (see \cref{fig:124Xe-crossover}).

The three rows show three interference regimes: (a) $\bmu$ is exactly degenerate ($|\rho_{ij}|=1$) with $\bsig$ or $\btau$, leading to sets of parallel lines, (b) $\beeta$ interferes at intermediate strength, and (c) $\blam$ barely interferes. Notice the intricate nature of where the combination of $0\nu\mm$ and $0\nu\pec$ is most useful. For both $\blam-\bsig$ and $\beeta-\btau$, the combined constraints are more powerful than for $\blam-\btau$ and $\beeta-\bsig$, respectively, because positron and electron mode constraints have opposite slopes. However, we stress that this plot is illustrative rather than a projection---the
$\pec$ half-lives shown are the crossover targets ($10^{27}$--$10^{29}$~yr), which far
exceed the NuDoubt$^{++}$ projection of $\sim\!10^{24}$~yr \cite{NuDoubt:2024jax}. 

\subsubsection*{Generalizing the Ratio Test} 
% \label{sec:general}
The single-current ratios of \cref{sec:DPratios} are only as strong as the assumption that one current dominates. In the following, we relax that assumption and demonstrate that the key discriminant, an anomalously small ratio, remains an unambiguous signature of $\blam$ even when several currents contribute simultaneously.

When several currents are considered at once, the ratio $\mathcal{R}$ is no longer one of the five fixed numbers of \cref{eq:DPratios} but depends on the actual admixture. Writing the two rates as the quadratic forms
with $c=(\bmu,\beeta,\blam,\btau,\bsig)$ being the vector of active couplings, their ratio is
\begin{equation}
\mathcal{R}(c) \;=\; \frac{T^{\pec}}{T^{\mm}} \;=\; \frac{c^{\mathsf T}\,C^{\mm}\,c}
                {c^{\mathsf T}\,C^{\pec}\,c}\,,
\label{eq:Rmix}
\end{equation}
which depends only on the \emph{direction} of $c$, as the overall coupling strength cancels. Hereby, $c$ ranges over all admixtures and one obtains an interval $[\mathcal{R}_{\min},\mathcal{R}_{\max}]$, whose endpoints are the smallest and largest \emph{generalized eigenvalues} $\lambda$ of the matrix pair $(C^{\mm},C^{\pec})$. A single active current, $c$, recovers the fixed value $\mathcal{R}_i=C^{\mm}_{i}/C^{\pec}_{i}$ of \cref{eq:DPratios}.

\begin{table}[t]
\centering
\caption{Range of $\mathcal{R}$ for every subset of the five currents, given by their corresponding extreme generalised eigenvalues. Of the 31 subsets, 16 are degenerate. Pairs are listed in the order of \cref{fig:rho-atlas}, whose four groups are reproduced here. All quadruples and the full set are degenerate and are omitted.}
\label{tab:subset-ranges}
\smallskip
\renewcommand{\arraystretch}{1.2}
\begin{tabular}{@{}llrr@{}}
\toprule
 & Subset & $\mathcal{R}_{\min}$ & $\mathcal{R}_{\max}$ \\
\midrule
\multirow{5}{*}{\emph{single currents}}
  & $\cmu$   & \multicolumn{2}{c}{199.2} \\
  & $\ceta$  & \multicolumn{2}{c}{306.6} \\
  & $\clam$  & \multicolumn{2}{c}{9.8} \\
  & $\ctau$  & \multicolumn{2}{c}{261.0} \\
  & $\csig$  & \multicolumn{2}{c}{94.5} \\
\midrule
\multirow{3}{*}{\emph{pairs: flat direction}}
  & $\cmu$--$\ctau$  & --- & --- \\
  & $\cmu$--$\csig$  & --- & --- \\
  & $\ctau$--$\csig$ & --- & --- \\
\midrule
\multirow{3}{*}{\emph{pairs: tilt flips}}
  & $\cmu$--$\ceta$  & 82.9 & 997.9 \\
  & $\ceta$--$\ctau$ & 96.7 & 1121.3 \\
  & $\ceta$--$\csig$ & 93.4 & 420.2 \\
\midrule
\multirow{3}{*}{\emph{pairs: weak tilt}}
  & $\cmu$--$\clam$  & 9.3 & 213.9 \\
  & $\clam$--$\ctau$ & 9.3 & 279.3 \\
  & $\clam$--$\csig$ & 9.7 & 97.8 \\
\midrule
\emph{pairs: decoupling}
  & $\ceta$--$\clam$ & 9.8 & 306.8 \\
\midrule
\multirow{3}{*}{\emph{non-degenerate triples}}
  & $\cmu$+$\ceta$+$\clam$  & 9.2 & 1053.9 \\
  & $\ceta$+$\clam$+$\ctau$ & 9.2 & 1188.4 \\
  & $\ceta$+$\clam$+$\csig$ & 9.7 & 427.8 \\
\bottomrule
\end{tabular}
\end{table}

We evaluate this interval for every two-current admixture---each of the ten current pairs, with the remaining three couplings set to zero, see \cref{tab:subset-ranges}. Allowing a second coupling to be present turns each single-current prediction into a range. Three of the ten pairs are drawn entirely from the exactly degenerate triplet of family~(a) ($\bmu, \btau, \bsig$), where $C^{\mm}$ has rank one. The remaining seven pairs give bounded intervals spanning $\mathcal{R}\in[\,9.3,\,1121\,]$. The upper limit is reached by having the two effective couplings $(\btau,\beeta)$ present at the same time, well above the single-current maximum $\mathcal{R}_{\beeta}\simeq307$ (in the hypothetical case of $\mathcal{R}_i > 307$, we would know that there is more than one current contributing, although this is surely in the experimentally inaccessible regime). The lower limit is obtained by having the two effective couplings $(\blam,\beeta)$ present at a time. 

The three combinations that do not contain $\blam$ have a lower-limit at $\mathcal{R}_{\min}=82.9$, $93.4$ and $96.7$ for $(\bmu,\beeta)$, $(\bsig,\beeta)$ and $(\btau,\beeta)$ respectively. All combinations that include $\blam$, however, feature a minimal value around $9.3$ to $9.8$, see \cref{tab:subset-ranges}. A measured ratio $\mathcal{R}\lesssim83$ would therefore establish that $\blam$ contributes, provided the signal is not caused by two or more members of the ($A_\nu$) triplet---the only configuration in which the ratio test loses its discriminating power altogether. This also strengthens our conclusion from the previous section, where our single current analysis pointed to a ratio $\mathcal{R}_i<95$. While these are large enough hierarchies for discrimination, we stress that the nuclear theory inputs still have a large uncertainty, which is the basis for intense theoretical effort, whose impact we discuss in~\cref{app:depend}.

It is natural to ask what range $\mathcal{R}$ can span when all five currents are active. Sixteen of the twenty-six subsets of size two or larger contain at least two members of the family~(a) triplet and inherit this behaviour. Meaningful bounds therefore exist with subsets containing at most one such member; these comprise $\beeta$, $\blam$ and one $A_\nu$ current and so have at
most three elements, for which
\begin{equation}
\mathcal{R}\;\in\;[\,9.2,\ 1188\,]\,,
\label{eq:fullrange}
\end{equation}
attained by $(\beeta,\blam,\btau)$, see \cref{tab:subset-ranges}. The pairwise values quoted above, $[9.3,1121]$, already approach both endpoints, so the two-current atlas of~\cref{fig:rho-atlas} is quantitatively representative of the general case wherever the ratio test retains any content at all.

A dedicated analysis should also consider the residual dependence of the uncertainties in the NME calculations. While we would like to point out this limitation, we leave this for further work, see~\cref{app:depend}.

%=============================================================================
\section{Relation to other works}
\label{sec:gks}
%=============================================================================
This analysis partially overlaps with the recent study of \cite{GKS26}, which likewise targets the positron modes of the NuDoubt$^{++}$ isotopes. Where the two analyses are degenerate, they agree, and we reproduce the results of \cite{GKS26}.

The essential difference is ---besides the scientific questions answered in this work (see \cref{sec:intro})---  the scale at which each study is performed. \cite{GKS26} quotes limits on the dimension-seven SMEFT operators defined at the new-physics scale $\Lambda_{\rm NP}$. Connecting that scale to a decay rate requires one-loop
renormalization-group running and matching down to $\Lambda_\chi\approx2$~GeV~\cite{Zhang2023,Zhang2024,GrafHati2025}. Along the way, several operators mix into the dimension-five Weinberg operator, whose chirally enhanced mass mechanism can then dominate the rate. 

In this work, we directly focus on the five effective couplings
$(\bmu,\beeta,\blam,\btau,\bsig)$ at the scale $\Lambda_\chi$. 
Considering this low energy basis, we do not need to consider running and matching effects.

The two analyses thus complement each other and converge to the same identified target. An operator whose low-energy footprint is generated primarily by loop mixing into $\mbb$ looks $\bmu$-like in every mode (see Footnote~\ref{footnote5}). Here, we study the genuinely non-$\bmu$ currents, above all the right-handed current $\blam$. In particular, we highlight the discriminating power of ratios across modes, previously not considered. 

%=============================================================================
\section{Conclusions}
\label{sec:conclusions}
%=============================================================================
Motivated by the proposed NuDoubt$^{++}$ experiment, we evaluate neutrinoless weak decay modes which emit positrons, $0\nu\pec$ and $0\nu\pp$, alongside the established $0\nu\mm$ decay mode. We put the positron modes in the context of the same EFT master formula developed in~\cite{Cirigliano2017, Cirigliano2018} in order to make all five long-range currents $(\bmu,\beeta,\blam,\btau,\bsig)$ directly comparable across the different modes. 

As expected, at equal half-life the positron modes are not competitive with $0\nu\mm$ in sensitivity, such that NuDoubt$^{++}$ cannot compete in sensitivity. Under the assumption of a single dominant current, however, an observed $0\nu\mm$ signal predicts a definite $0\nu\pec$ half-life, as discussed in \cref{sec:crossing}. While the right-handed current, $\blam$, is reached first, every other channel requires additional orders of magnitude in sensitivity. Despite this, due to the large hierarchy between the predicted $0\nu\pec$ half-life for $\blam$ with respect to others, the positron programme of NuDoubt$^{++}$, in particular $0\nu\pec$ can be used to help \emph{pinpoint} the mechanism of the neutrinoless weak decay.

To formalize this, we introduce half-life ratios in \cref{sec:DPratios}, demonstrating that  in particular half-life ratios across different modes---in contrast to previous proposals of taking only ratios of different $0\nu\mm$ isotopes~\cite{Deppisch:2006hb}---can give discriminating power. 

This is true even for a $0\nu\pec$ \emph{non}-observation, which already constrains the $\blam$ contribution to an observed $0\nu\mm$ signal, leading to an upper limit on the $\blam$ coupling strength. 

While we neglected interferences as a first approximation at this stage, we later highlighted that this is actually a good approximation as $\blam$ interferes only weakly, especially for $0\nu\mm$ and $0\nu\pec$, as can be noted from~\cref{fig:rho-atlas}.

Finally, in \cref{sec:interference}, we discussed the discriminating power across modes when including interferences. For example, the smallest $\mathcal{R}$ reachable without $\blam$ reduces only slightly from $\mathcal{R}_{\bsig}=94.5$ to $82.9$ once interference is included. Furthermore,
% more than one current at a time
we showed that there is an \emph{exact} degeneracy between $\bmu, \btau$ and $\bsig$ and a limit cannot be set on one of these currents, if several of them contribute at the same time. However, in many other cases, the positron mode $0\nu\pec$, which has a different phenomenology with respect to the modes with two continuum leptons, can break partial degeneracies (in contrast to considering only $0\nu \mm$ in different isotopes). For instance, a measured ratio $\mathcal{R}\lesssim83$ would point towards a contribution of $\blam$, provided the signal is not caused by two or more members of the ($\bmu, \btau$ and $\bsig$) triplet.

While the positron modes will not race $0\nu\mm$ experiments to discovery; they could help us in understanding what kind of physics a discovery implies.
% ======================================================
\begin{acknowledgments}
\noindent
The authors gratefully acknowledge useful discussions with Martin Hirsch, Jenni Kotila, Ovidiu Nițescu and Fedor Šimkovic. We also thank Michael Wurm and Stefan Schoppmann for discussions and support, and the wider NuDoubt$^{++}$ Collaboration for their interest in this work. JH and GAP are supported by the Cluster of Excellence \emph{Precision Physics, Fundamental Interactions and Structure of Matter} (PRISMA++, EXC 2118/2, Project ID 390831469), funded by the German Research Foundation (DFG).
\end{acknowledgments}
%=============================================================================
% \newpage
\bibliography{refs}
%=============================================================================
\appendix
%=============================================================================

\section{INPUT MAPPING:\\ DKT in the EFT basis}
\label{app:dkt}

In this section, we discuss the various nuclear inputs. Most importantly, we translate the Doi--Kotani--Takasugi matrix-element combinations of~\cite{Muto1989,Hirsch1994} into the EFT sub-amplitude basis of \cref{eq:amps} \cite{Cirigliano2017,Cirigliano2018}, so that the legacy nuclear inputs can be used within the master formula. We will then show why that means that some currents can be partially constrained from the legacy data, like $\bsig$, and some cannot be constrained at all, like $\btau$. Finally, we will demonstrate how changing the $0\nu\mm$ nuclear data inputs (which is currently~\cite{Horoi2017}) can affect our results.

Throughout this appendix ``DKT'' denotes the Doi--Kotani--Takasugi parametrisation of the neutrinoless weak decay rate~\cite{Doi:1985dx,Doi:1992dm}, the standard language of the pnQRPA literature and the form in which~\cite{Muto1989,Hirsch1994} tabulate their results. It organises the rate around the combinations $M_{1\pm}$, $M_{2\pm}$ of \cref{eq:M12}, together with
$M_P$ and the recoil element $M_R$; the correspondence with the coupling basis used here is collected in~\cref{tab:conventions}. Two structures of the EFT
decomposition have no DKT counterpart: the axial--pseudoscalar (AP) and pseudoscalar--pseudoscalar (PP) spin--isospin matrix elements and the short-distance Fermi element $M_{F,\rm sd}$. This is what limits the
reinterpretation reach discussed in \cref{app:other-currents}.

\subsection{Matrix-element mapping}
\label{app:mapping}

Defining~\cite{Muto1989}
\begin{align}
M_{1\pm} &= M_{GTq} \pm 3M_{Fq} - 6M_T\,, \notag\\
M_{2\pm} &= M_{GT\omega} \pm M_{F\omega} - \tfrac{1}{9}M_{1\mp}\,,
\label{eq:M12}
\end{align}
matching the $G_{02}$, $G_{03}$, $G_{04}$ and $G_{09}$ coefficients gives
\begin{align}
M_{E,L} &= -\tfrac{1}{4}V_{ud}M_{2+}\,, &
M_{E,R} &= +\tfrac{1}{4}V_{ud}M_{2-}\,, \notag\\
M_{m_e,L} &= -\tfrac{1}{18}V_{ud}M_{1-}\,, &
\mathcal{M}_M &= \frac{V_{ud}}{R_A m_N}M_R\,.
\label{eq:direct}
\end{align}
The relative sign of $M_{E,L}$ and $M_{E,R}$ is fixed by requiring the correct sign of $C_{\beeta\blam}$. Since the DKT convention absorbs $(g_V/g_A)^2$ into the Fermi element~\cite{Doi:1985dx,Muto1989}, the EFT-basis value is $M_F^{\rm(EFT)}=(g_A/g_V)^2M_F^{\rm(DKT)}$.

\subsection{Sub-leading corrections}
\label{app:subleading}

\Cref{eq:master} omits three terms, sub-leading in the chiral counting~\cite{Cirigliano2018} but relevant at the few-percent level, which we restore additively: the pseudoscalar--recoil cross-terms
$M_P^2G_8-S_eM_PM_RG_7$ in the diagonal $C_{\beeta\beeta}$; the pseudoscalar
piece $-S_e(M_{GT}-M_F)M_PG_5$ in $C_{\bmu\beeta}$; and a structural difference in $C_{\beeta\blam}$, where the EFT amplitude forces the $G_{03}$ cross-term into
$(M_{1-}M_{2-}+M_{1+}M_{2+})$ while DKT give $(M_{1+}M_{2-}+M_{1-}M_{2+})$, the difference $\propto(M_{1+}-M_{1-})(M_{2+}-M_{2-})$ being added explicitly. These follow from matching the two rate expressions term by term.

\subsection{Numerical validation}
\label{app:validation}

We have verified the implementation against DKT for seven $\mm$ isotopes using
the NMEs of Ref.~\cite{Muto1989} with the PSFs of~\cite{Doi:1985dx}, and for
four $0\nu\pec$ isotopes ($^{58}$Ni, $^{96}$Ru, $^{106}$Cd, $^{124}$Xe) using~\cite{Hirsch1994}. All single-current limits and interference coefficients agree to double precision once the corrections of \cref{app:subleading} are included. These legacy inputs serve for validation only.
% \hspace{-0.5cm}
\subsection{Tensor reach from \texorpdfstring{$M_R$}{MR}}
\label{app:other-currents}

\begin{table}[b]
\centering
\caption{Tensor limits from the legacy tables at the 1989/1994 half-lives:
magnetic-only (from $M_R$) versus $M_{F,\rm sd}$ restored by the chiral
estimate $|M_{F,\rm sd}|\sim|M_F|$, at $g_{T'}=1$.}
\label{tab:tensor-partial-full}
\smallskip
\begin{tabular}{@{}lccc@{}}
\toprule
Isotope & $\btau$ partial & $\btau$ full & ratio \\
\midrule
$^{76}$Ge  & $1.07\times10^{-8}$ & $9.61\times10^{-9}$ & 0.897 \\
$^{82}$Se  & $3.68\times10^{-8}$ & $3.30\times10^{-8}$ & 0.896 \\
$^{100}$Mo & $1.14\times10^{-7}$ & $1.02\times10^{-7}$ & 0.896 \\
$^{128}$Te & $6.02\times10^{-9}$ & $5.40\times10^{-9}$ & 0.898 \\
$^{130}$Te & $8.29\times10^{-8}$ & $7.47\times10^{-8}$ & 0.901 \\
$^{136}$Xe & $1.30\times10^{-7}$ & $1.20\times10^{-7}$ & 0.918 \\
$^{150}$Nd & $1.72\times10^{-8}$ & $1.55\times10^{-8}$ & 0.897 \\
\midrule
$^{58}$Ni  & $1.47\times10^{-4}$ & $1.33\times10^{-4}$ & 0.905 \\
$^{96}$Ru  & $5.22\times10^{-4}$ & $4.68\times10^{-4}$ & 0.897 \\
$^{106}$Cd & $1.25\times10^{-4}$ & $1.12\times10^{-4}$ & 0.895 \\
$^{124}$Xe & $6.31\times10^{-5}$ & $5.64\times10^{-5}$ & 0.894 \\
\bottomrule
\end{tabular}
\label{tab:tensor}
\end{table}

Of the two currents beyond the DKT basis, $\btau$ is partially accessible and $\bsig$ is not. The tensor amplitude carries
\begin{equation}
M_{T6} = \frac{2g_{T'}g_V}{g_Tg_A^2}\frac{m_\pi^2}{m_N^2}M_{F,\rm sd}
       - \frac{8}{g_M}\bigl(M_{GT}^{MM}+M_T^{MM}\bigr),
\label{eq:MT6}
\end{equation}
whose second piece follows from $M_R$ via
$M_{GT}^{MM}+M_T^{MM}=(g_M/2g_A)\mathcal{M}_M$. The first is inaccessible:
$M_{F,\rm sd}$ is not tabulated in~\cite{Muto1989,Hirsch1994}. Its effect is bounded by the chiral prefactor $(m_\pi/m_N)^2=2.21\times10^{-2}$, or
$2.76\times10^{-2}$ including the coupling factor at $g_{T'}=1$. Restoring it with $|M_{F,\rm sd}|\sim|M_F|$ enhances $|M_{T6}|$ coherently and tightens the
limit by a $90\%$ factor ($0.894$--$0.918$) across all eleven isotopes, see
\cref{tab:tensor-partial-full}. The dominant residual uncertainty is the \emph{sign} of $M_{F,\rm sd}$: the opposite sign relaxes the limit by a comparable amount, so \cref{tab:tensor-partial-full} sets the scale of the
shift while presenting the lower possibility.

Note the effect in~\cref{tab:tensor} depends on the unknown LEC $g_{T'}$, which is quoted
at $g_{T'}=1$ in ~\cref{tab:tensor}, while set to zero in our work in accordance to the conservative choice of \cite{GKS26}. At $g_{T'}=0$ the $M_{F,\rm sd}$ term would vanish identically and every ratio in \cref{tab:tensor-partial-full} would be exactly unity. Therefore, the results in~\cref{tab:tensor} should be interpreted as the maximal size of the missing piece rather than as a correction applied anywhere in the main results.

\subsection{Why \texorpdfstring{$\bsig$}{sigma} cannot be bounded from DKT}
\label{app:depend}

The scalar/pseudoscalar current enters through
\begin{equation}
A_\nu \supset V_{ud}\,\frac{B_\chi}{m_e}\,(C_{SL}-C_{SR})\,M_{PS},
\label{eq:Anu-scalar}
\end{equation}
with $M_{PS}=\tfrac12M_{GT}^{AP}+M_{GT}^{PP}+\tfrac12M_T^{AP}+M_T^{PP}$ and $B_\chi=m_\pi^2/(m_u+m_d)$. No AP/PP piece appears in the DKT decomposition, so $M_{PS}$ vanishes identically on DKT inputs and there is no $\bsig$ constraint at all. Bounding $\bsig$ requires a modern decomposition: the shell-model NMEs of~\cite{Horoi2017}, the tables of~\cite{Cirigliano2017}, or the IBM-2 tables of~\cite{Barea2015, GKS26}.

Note, the scalar effective coupling is dependent on $M_{PS}$ contributing to the sub-amplitude $A_\nu$ of \cref{eq:amps}. In ~\cite{Cirigliano2017}, it was observed that the shell-model NMEs~\cite{Horoi2017} carry the opposite sign to the QRPA and IBM-2 results it compares against, but with similar magnitudes. Although this is irrelevant for the single-coupling limits, once several operators are present this has an effect. As our $\mm$ inputs follow~\cite{Horoi2017}, this would only affect the $\bsig$ entries in~\cref{fig:rho-atlas}, in this case, the $\rho_{\bmu\bsig}$ and $\rho_{\btau\bsig}$ take the same
signs in both modes, and $\rho_{\beeta\bsig}$ would acquires the mode-dependent sign-difference shown by $\rho_{\bmu\beeta}$ and $\rho_{\beeta\btau}$. The structural conclusions are untouched, but the `$\blam$-free' threshold would then become set by
% $(\bsig,\beeta)$ 
at $\mathcal{R}\simeq50$ rather than 
% by $(\bmu,\beeta)$
at $\mathcal{R}\simeq83$.
While this decreases the hierarchy between the ratios with and without $\blam$, it still features a sizeable ratio difference for discrimination.

\end{document}